\pdfoutput=1
\documentclass[12pt,a4paper]{article}

\usepackage{ifthen} 
\newboolean{pdflatex}
\setboolean{pdflatex}{true} 

\newboolean{articletitles}
\setboolean{articletitles}{true} 

\newboolean{uprightparticles}
\setboolean{uprightparticles}{false} 

\newboolean{inbibliography}
\setboolean{inbibliography}{false} 

\usepackage{microtype}
\usepackage{lineno}  
\usepackage{xspace} 

\usepackage{graphicx}  
\usepackage{color}
\usepackage{colortbl}
\graphicspath{{./figs/}} 

\usepackage{amsmath} 
\usepackage{amssymb}
\usepackage{amsfonts}
\usepackage{upgreek} 

\newcommand*\patchAmsMathEnvironmentForLineno[1]{%
\expandafter\let\csname old#1\expandafter\endcsname\csname #1\endcsname
\expandafter\let\csname oldend#1\expandafter\endcsname\csname
end#1\endcsname
 \renewenvironment{#1}%
   {\linenomath\csname old#1\endcsname}%
   {\csname oldend#1\endcsname\endlinenomath}%
}
\newcommand*\patchBothAmsMathEnvironmentsForLineno[1]{%
  \patchAmsMathEnvironmentForLineno{#1}%
  \patchAmsMathEnvironmentForLineno{#1*}%
}
\AtBeginDocument{%
\patchBothAmsMathEnvironmentsForLineno{equation}%
\patchBothAmsMathEnvironmentsForLineno{align}%
\patchBothAmsMathEnvironmentsForLineno{flalign}%
\patchBothAmsMathEnvironmentsForLineno{alignat}%
\patchBothAmsMathEnvironmentsForLineno{gather}%
\patchBothAmsMathEnvironmentsForLineno{multline}%
}

\usepackage{hyperref}    
\usepackage[all]{hypcap} 

\def\lhcb {\mbox{LHCb}\xspace}

\def\cleo   {\mbox{CLEO}\xspace}

\def\rich   {RICH\xspace}

\ifthenelse{\boolean{uprightparticles}}%
{

 \def\Ppi         {\ensuremath{\uppi}\xspace}

 \def\PDelta      {\ensuremath{\Delta}\xspace}                 
 \def\PXi      {\ensuremath{\Xi}\xspace}                 
 \def\PLambda      {\ensuremath{\Lambda}\xspace}                 
 \def\PSigma      {\ensuremath{\Sigma}\xspace}                 
 \def\POmega      {\ensuremath{\Omega}\xspace}                 
 \def\PUpsilon      {\ensuremath{\Upsilon}\xspace}                 
 
 \def\PB      {\ensuremath{\mathrm{B}}\xspace}                 
                  
 \def\PD      {\ensuremath{\mathrm{D}}\xspace}

 \def\PK      {\ensuremath{\mathrm{K}}\xspace}

 \def\Pb      {\ensuremath{\mathrm{b}}\xspace}                 
 \def\Pc      {\ensuremath{\mathrm{c}}\xspace}

 \def\Pi      {\ensuremath{\mathrm{i}}\xspace}

 \def\Pp      {\ensuremath{\mathrm{p}}\xspace}

}
{

 \def\Ppi         {\ensuremath{\pi}\xspace}

 \mathchardef\PDelta="7101
 \mathchardef\PXi="7104
 \mathchardef\PLambda="7103
 \mathchardef\PSigma="7106
 \mathchardef\POmega="710A
 \mathchardef\PUpsilon="7107
                  
 \def\PB      {\ensuremath{B}\xspace}                 
                  
 \def\PD      {\ensuremath{D}\xspace}

 \def\PK      {\ensuremath{K}\xspace}

 \def\Pb      {\ensuremath{b}\xspace}                 
 \def\Pc      {\ensuremath{c}\xspace}

 \def\Pi      {\ensuremath{i}\xspace}

 \def\Pp      {\ensuremath{p}\xspace}

}

\def\cquark    {\ensuremath{\Pc}\xspace}

\def\bquark    {\ensuremath{\Pb}\xspace}

\def\pion  {\ensuremath{\Ppi}\xspace}
\def\piz   {\ensuremath{\pion^0}\xspace}

\def\pip   {\ensuremath{\pion^+}\xspace}
\def\pim   {\ensuremath{\pion^-}\xspace}

\def\kaon  {\ensuremath{\PK}\xspace}
  \def\Kbar  {\kern 0.2em\overline{\kern -0.2em \PK}{}\xspace}

\def\Kp    {\ensuremath{\kaon^+}\xspace}
\def\Km    {\ensuremath{\kaon^-}\xspace}

  \def\Dbar    {\kern 0.2em\overline{\kern -0.2em \PD}{}\xspace}
\def\D       {\ensuremath{\PD}\xspace}

\def\Dz      {\ensuremath{\D^0}\xspace}
\def\Dzb     {\ensuremath{\Dbar^0}\xspace}
\def\Dstar   {\ensuremath{\D^*}\xspace}

\def\Dstarp  {\ensuremath{\D^{*+}}\xspace}
\def\Dstarm  {\ensuremath{\D^{*-}}\xspace}

\def\Bbar    {\ensuremath{\kern 0.18em\overline{\kern -0.18em \PB}{}}\xspace}

  \def\Y#1S{\ensuremath{\PUpsilon{(#1S)}}\xspace}

\def\proton      {\ensuremath{\Pp}\xspace}

\def\Lbar {\ensuremath{\kern 0.1em\overline{\kern -0.1em\PLambda}}\xspace}

\newcommand{\decay}[2]{\ensuremath{#1\!\to #2}\xspace}         

\def\to                 {\ensuremath{\rightarrow}\xspace}

\def\CP                {\ensuremath{C\!P}\xspace}

\def\AT#1     {\ensuremath{A_{\mathrm{T}}^{#1}}\xspace}           

\def\C#1      {\ensuremath{\mathcal{C}_{#1}}\xspace}                       
\def\Cp#1     {\ensuremath{\mathcal{C}_{#1}^{'}}\xspace}                    
\def\Ceff#1   {\ensuremath{\mathcal{C}_{#1}^{\mathrm{(eff)}}}\xspace}        
\def\Cpeff#1  {\ensuremath{\mathcal{C}_{#1}^{'\mathrm{(eff)}}}\xspace}       
\def\Ope#1    {\ensuremath{\mathcal{O}_{#1}}\xspace}                       
\def\Opep#1   {\ensuremath{\mathcal{O}_{#1}^{'}}\xspace}                    

\newcommand{\tev}{\ifthenelse{\boolean{inbibliography}}{\ensuremath{~T\kern -0.05em eV}\xspace}{\ensuremath{\mathrm{\,Te\kern -0.1em V}}\xspace}}
\newcommand{\gev}{\ensuremath{\mathrm{\,Ge\kern -0.1em V}}\xspace}
\newcommand{\mev}{\ensuremath{\mathrm{\,Me\kern -0.1em V}}\xspace}
\newcommand{\kev}{\ensuremath{\mathrm{\,ke\kern -0.1em V}}\xspace}
\newcommand{\ev}{\ensuremath{\mathrm{\,e\kern -0.1em V}}\xspace}
\newcommand{\gevc}{\ensuremath{{\mathrm{\,Ge\kern -0.1em V\!/}c}}\xspace}
\newcommand{\mevc}{\ensuremath{{\mathrm{\,Me\kern -0.1em V\!/}c}}\xspace}
\newcommand{\gevcc}{\ensuremath{{\mathrm{\,Ge\kern -0.1em V\!/}c^2}}\xspace}
\newcommand{\gevgevcccc}{\ensuremath{{\mathrm{\,Ge\kern -0.1em V^2\!/}c^4}}\xspace}
\newcommand{\mevcc}{\ensuremath{{\mathrm{\,Me\kern -0.1em V\!/}c^2}}\xspace}

\def\mm   {\ensuremath{\rm \,mm}\xspace}

\def\mum  {\ensuremath{\,\upmu\rm m}\xspace}

\def\invfb   {\ensuremath{\mbox{\,fb}^{-1}}\xspace}

\newcommand{\chisq}{\ensuremath{\chi^2}\xspace}
\newcommand{\chisqndf}{\ensuremath{\chi^2/\mathrm{ndf}}\xspace}
\newcommand{\chisqip}{\ensuremath{\chi^2_{\rm IP}}\xspace}

\def\gsim{{~\raise.15em\hbox{$>$}\kern-.85em
          \lower.35em\hbox{$\sim$}~}\xspace}
\def\lsim{{~\raise.15em\hbox{$<$}\kern-.85em
          \lower.35em\hbox{$\sim$}~}\xspace}

\def\sPlot{\mbox{\em sPlot}}

\def\ptot       {\mbox{$p$}\xspace}
\def\pt         {\mbox{$p_{\rm T}$}\xspace}

\def\tell1  {TELL1\xspace}
\def\ukl1   {UKL1\xspace}

\usepackage{cite} 
\usepackage{mciteplus}

\newcommand{\figs}[1]{Figs.~\ref{#1}} 
\newcommand{\fig}[1]{Fig.~\ref{#1}} 
\newcommand{\Fig}[1]{Figure~\ref{#1}} 
\newcommand{\tab}[1]{Table~\ref{#1}} 
\newcommand{\eqn}[1]{Eq.~\ref{#1}} 

\def\DFourPi{\mbox{\ensuremath{\decay{\Dz}{\pim \pip \pip \pim}}}\xspace}
\def\DKKPiPi{\mbox{\ensuremath{\decay{\Dz}{\Km \Kp \pim \pip}}}\xspace} 
\def\DKThreePi{\mbox{\ensuremath{\decay{\Dz}{\Km \pip \pip \pim}}}\xspace}
 
\def\FourPi {\ensuremath{\pim \pip \pip \pim}\xspace}
\def\KKPiPi {\ensuremath{\Km \Kp \pim \pip}\xspace}
\def\KThreePi {\ensuremath{\Km \pip \pip \pim}\xspace}
\def\DsKKPiPiPi{\ensuremath{\decay{\D_{s}^{+}}{\Km \Kp \pim \pip \pip}}\xspace}

\def\DKPiPiPiPiPiZ{\ensuremath{\decay{\Dz}{\Km \pip \pim \pip \piz}}\xspace}

\newcommand{\SCP}{\ensuremath{S_{\CP}}\xspace}
\newcommand{\fourBody}{four-body\xspace}
\newcommand{\threeBody}{three-body\xspace}

\def\pvalue {$p$-value\xspace}
\def\pvalues {$p$-values\xspace}
\def\deltam{\ensuremath{\Delta m}\xspace}

\def\mD{\ensuremath{m(hhhh)}\xspace}

\usepackage{graphicx}
\usepackage{caption}
\usepackage[lofdepth,lotdepth]{subfig}

\usepackage{flafter}
\usepackage{afterpage}
\begin{document}

\renewcommand{\thefootnote}{\fnsymbol{footnote}}
\setcounter{footnote}{1}


\begin{titlepage}
\pagenumbering{roman}

\vspace*{-1.5cm}
\centerline{\large EUROPEAN ORGANIZATION FOR NUCLEAR RESEARCH (CERN)}
\vspace*{1.5cm}
\hspace*{-0.5cm}
\begin{tabular*}{\linewidth}{lc@{\extracolsep{\fill}}r}
\ifthenelse{\boolean{pdflatex}}
{\vspace*{-2.7cm}\mbox{\!\!\!\includegraphics[width=.14\textwidth]{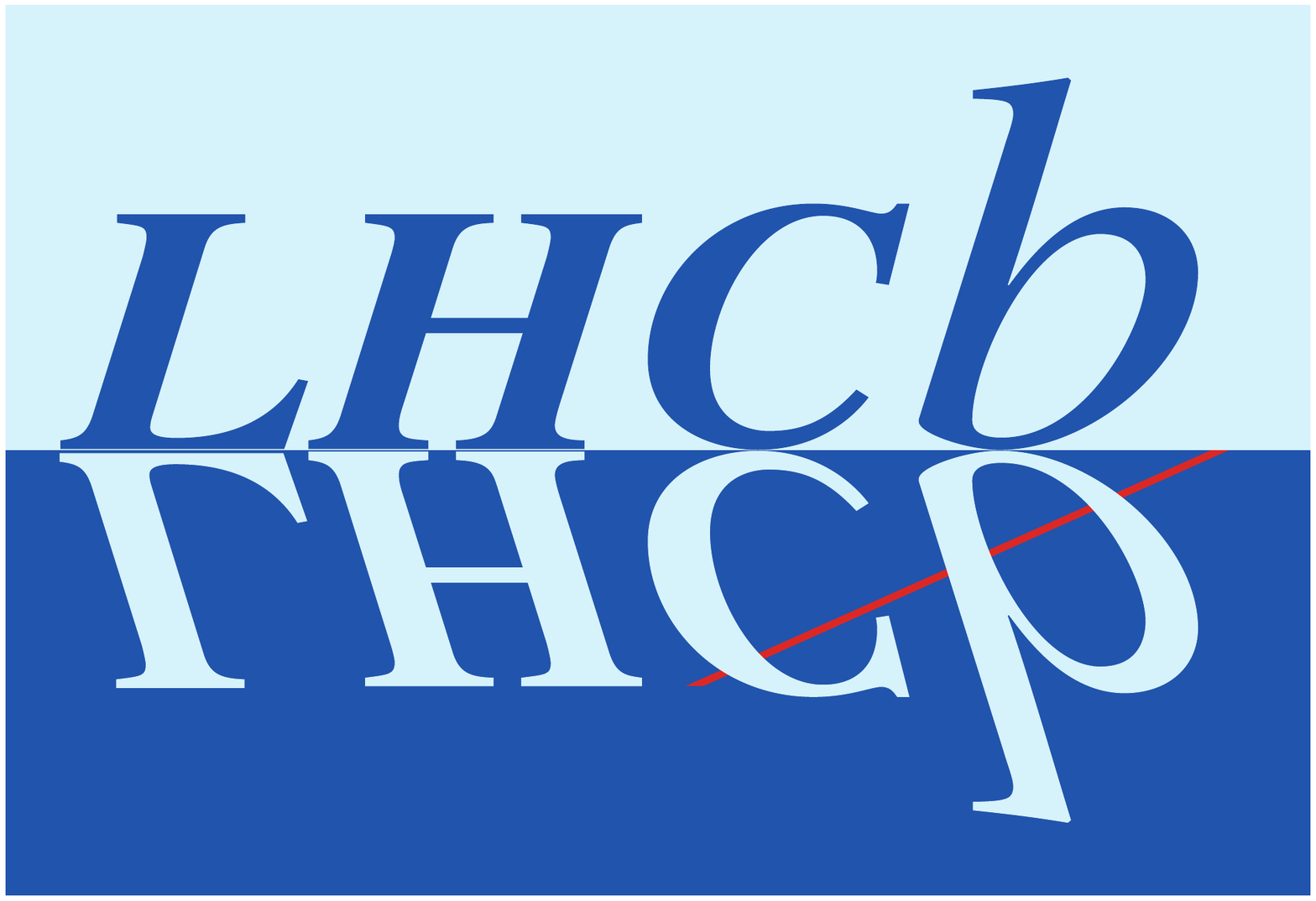}} & &}%
{\vspace*{-1.2cm}\mbox{\!\!\!\includegraphics[width=.12\textwidth]{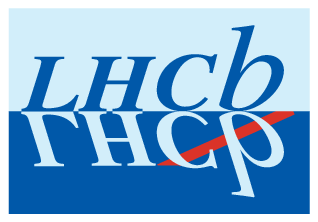}} & &}%
\\
 & & CERN-PH-EP-2013-151 \\  
 & & LHCb-PAPER-2013-041 \\  
 & & 15 August 2013 \\ 
 & & \\
\end{tabular*}

\vspace*{3.0cm}

{\bf\boldmath\huge
\begin{center}
  Model-independent search for \CP violation in \DKKPiPi and \DFourPi decays
\end{center}
}

\vspace*{2.0cm}

\begin{center}
The LHCb collaboration\footnote{Authors are listed on the following pages.}
\end{center}

\vspace{\fill}

\begin{abstract}
  \noindent
A search for \CP violation in the phase-space structures of \Dz and \Dzb decays to the final states \KKPiPi and \FourPi is presented.
The search is carried out with a data set corresponding to an integrated luminosity of 1.0\invfb 
collected in 2011 by the \lhcb experiment in $\proton\proton$ collisions at a centre-of-mass energy of 7\tev. 
For the \KKPiPi final state, the \fourBody phase space is divided into 32 bins, each bin with approximately 1800 decays. 
The \pvalue under the hypothesis of no \CP violation is 9.1\%, and in no bin is a \CP asymmetry greater than 6.5\% observed.
The phase space of the \FourPi final state is partitioned into 128 bins, each bin with approximately 2500 decays. The \pvalue under the hypothesis of no \CP violation is 41\%, 
and in no bin is a \CP asymmetry greater than 5.5\% observed. All results are consistent with the hypothesis of no \CP violation at the current sensitivity.

\end{abstract}

\vspace*{1.0cm}

\begin{center}
  Submitted to Phys.~Lett.~B 
\end{center}

\vspace{\fill}

{\footnotesize 
\centerline{\copyright~CERN on behalf of the \lhcb collaboration, license \href{http://creativecommons.org/licenses/by/3.0/}{CC-BY-3.0}.}}
\vspace*{2mm}

\end{titlepage}


\newpage
\setcounter{page}{2}
\mbox{~}
\newpage

\centerline{\large\bf LHCb collaboration}
\begin{flushleft}
\small
R.~Aaij$^{40}$, 
B.~Adeva$^{36}$, 
M.~Adinolfi$^{45}$, 
C.~Adrover$^{6}$, 
A.~Affolder$^{51}$, 
Z.~Ajaltouni$^{5}$, 
J.~Albrecht$^{9}$, 
F.~Alessio$^{37}$, 
M.~Alexander$^{50}$, 
S.~Ali$^{40}$
% ...

{\footnotesize \it
% ...
$ ^{5}$Clermont Universit\'{e}, Universit\'{e} Blaise Pascal, CNRS/IN2P3, LPC, Clermont-Ferrand, France\\
$ ^{6}$CPPM, Aix-Marseille Universit\'{e}, CNRS/IN2P3, Marseille, France\\
% ...
$ ^{9}$Fakult\"{a}t Physik, Technische Universit\"{a}t Dortmund, Dortmund, Germany\\
% ...
$ ^{36}$Universidad de Santiago de Compostela, Santiago de Compostela, Spain\\
$ ^{37}$European Organization for Nuclear Research (CERN), Geneva, Switzerland\\
% ...
$ ^{40}$Nikhef National Institute for Subatomic Physics, Amsterdam, The Netherlands\\
% ...
$ ^{45}$H.H. Wills Physics Laboratory, University of Bristol, Bristol, United Kingdom\\
% ...
$ ^{50}$School of Physics and Astronomy, University of Glasgow, Glasgow, United Kingdom\\
$ ^{51}$Oliver Lodge Laboratory, University of Liverpool, Liverpool, United Kingdom\\
% ...
}
\end{flushleft}

\cleardoublepage

\renewcommand{\thefootnote}{\arabic{footnote}}
\setcounter{footnote}{0}

\def\sweighted{s-weighted\xspace}
\def\SCPi{\ensuremath{S_{\CP}^{i}}\xspace}
\def\invariantMassCombs{s(1,2), s(2,3), s(1,2,3), s(2,3,4), and s(3,4)\xspace}
\def\invariantMassCombsKKPiPi{\mbox{s(\Km, \Kp)}, \mbox{s(\Kp, \pim)}, \mbox{s(\Km, \Kp, \pim)}, \mbox{s(\Kp, \pim, \pip)}, and \mbox{s(\pim, \pip)}\xspace}
\def\invariantMassCombsFourPi{\mbox{s(\pim, \pip)}, \mbox{s(\pip, \pip)}, \mbox{s(\pim, \pip, \pip)}, \mbox{s(\pip, \pip, \pim)}, and \mbox{s(\pip, \pim)}\xspace}

\pagestyle{plain} \setcounter{page}{1}
\pagenumbering{arabic}


\section{Introduction}
\label{sec:Introduction}

Standard Model predictions for the magnitude of \CP violation (CPV) in charm meson decays are generally of $\mathcal{O}$($10^{-3}$)~\cite{Bianco:2003vb,Du:2006jc}, although values up to $\mathcal{O}$($10^{-2}$) cannot be ruled out~\cite{Buccella:2013tya,Bobrowski:2010xg}. The size of CPV can be significantly enhanced in new physics models~\cite{Grossman:2006jg,Petrov:2010gy}, making charm transitions a promising area to search for new physics. 
Previous searches for CPV in charm decays caused a large interest
in the community~\cite{LHCb-PAPER-2013-003,LHCb-PAPER-2012-052,LHCb-PAPER-2011-023} and justify detailed searches for CPV in many different final states. 
Direct CPV can occur when at least two amplitudes interfere with strong and weak phases that each differ from one another. Singly-Cabibbo-suppressed charm hadron decays, where both tree processes and electroweak loop processes can contribute, are promising channels with which to search for CPV. The rich structure of interfering amplitudes makes \fourBody decays ideal to perform such searches. 

The phase-space structures of the \DKKPiPi and \DFourPi decays\footnote{Unless otherwise specified, inclusion of charge-conjugate processes is implied.} are investigated for localised CPV in a manner that is independent of an amplitude model of the \Dz meson decay. The Cabibbo-favoured \DKThreePi decay, where direct CPV can not occur in the Standard Model, is used as a control channel. A model-dependent search for CPV in \DKKPiPi was previously carried out by the \cleo collaboration~\cite{Artuso:2012df} with a data set of approximately 3000 signal decays, where no evidence for CPV was observed. This analysis is carried out on a data set of approximately $5.7\times10^{4}$ \DKKPiPi decays and $3.3\times10^{5}$ \DFourPi decays. The data set is based on an integrated luminosity of 1.0\invfb of $\proton\proton$ collisions
with a centre-of-mass energy of 7\tev,  recorded by the \lhcb experiment during 2011.
The analysis is based on  \Dz mesons produced in $\decay{\Dstarp}{\Dz \pi^{+}}$ decays. The charge of the soft pion ($\pi^{+}$) identifies the flavour of the meson at production. The phase space is partitioned into $N_{\rm bins}$  bins, and the significance of the difference in population between \CP conjugate decays for each bin is calculated as
\begin{equation}
  	        \SCPi = \frac{N_{i}(\Dz) - \alpha N_{i}(\Dzb)}{\sqrt{\alpha \left(  \sigma_{i}^{2}(\Dz) + \sigma_{i}^{2}(\Dzb) \right)} }, 	~~~	\alpha = \frac{\sum_{i}{N_{i}(\Dz)}}{\sum_{i}{N_{i}(\Dzb)}}, \label{eq:SCP}
\end{equation}
where $N_{i}$ is the number of signal decays in bin $i$, and $\sigma_{i}$ is the associated uncertainty in the number of signal decays in bin $i$~\cite{Bediaga:2009tr}. 
The normalisation constant $\alpha$ removes global production and detection differences between \Dstarp and \Dstarm decays.

In the absence of any asymmetry, \SCP is Gaussian distributed with a mean of zero and a width of one. A significant variation from a unit Gaussian distribution indicates the presence of an asymmetry.
The sum of squared \SCP values is a \chisq statistic, 
\begin{equation*}
\chisq = \sum_{i}{ \left( \SCPi \right)^{2}},
\end{equation*}
with $N_{\mathrm{bins}} - 1$ degrees of freedom, from which a \pvalue is calculated. Previous analyses of \threeBody \PD meson decays have employed similar analysis techniques~\cite{Aubert:2008yd,LHCb-PAPER-2011-017}.

\section{Detector}

The \lhcb detector~\cite{Alves:2008zz} is a single-arm forward
spectrometer covering the \mbox{pseudorapidity} range $2<\eta <5$,
designed for the study of particles containing \bquark or \cquark
quarks. The detector includes a high-precision tracking system
consisting of a silicon-strip vertex detector surrounding the $pp$
interaction region, a large-area silicon-strip detector located
upstream of a dipole magnet with a vertically oriented magnetic field and bending power of about
$4{\rm\,Tm}$, and three stations of silicon-strip detectors and straw
drift tubes placed downstream. To alleviate the impact of charged particle-antiparticle detection asymmetries,
the magnetic field polarity is switched regularly, and data are taken in each polarity. 
The two magnet polarities are henceforth referred to as ``magnet up'' and ``magnet down''.
The combined tracking system provides momentum measurement with
relative uncertainty that varies from 0.4\% at 5\gevc to 0.6\% at 100\gevc,
and impact parameter resolution of 20\mum for
tracks with high transverse momentum. Charged hadrons are identified
with two ring-imaging Cherenkov (\rich) detectors~\cite{LHCb-DP-2012-003}. Photon, electron, and
hadron candidates are identified by a calorimeter system consisting of
scintillating-pad and preshower detectors, an electromagnetic
calorimeter, and a hadronic calorimeter. Muons are identified by a
system composed of alternating layers of iron and multiwire
proportional chambers.
The trigger consists of a hardware stage, based on information from the calorimeter and muon
systems, followed by a software stage~\cite{LHCb-DP-2012-004}. Events are required to pass both hardware and software trigger levels. 
The software trigger optimised for the reconstruction of \fourBody 
hadronic charm decays requires a four-track secondary vertex with a scalar sum of the transverse 
momenta, \pt, of the tracks greater than $2\gevc$. At least two tracks are required to have 
$\pt > 500\mevc$ and momentum, \ptot, greater than $5\gevc$. The remaining two 
tracks are required to have $\pt > 250\mevc$ and $\ptot > 2\gevc$. 
A requirement is also imposed on the \chisq of the impact parameter (\chisqip) of the remaining two 
tracks with respect to any primary interaction to be greater than 10, where \chisqip is defined as the
difference in \chisq of a given primary vertex reconstructed with and
without the considered track. 

\section{Selection}

Candidate \Dz decays are reconstructed from combinations of pion and kaon 
candidate tracks.
The \Dz candidates are required to have $\pt >  3\gevc$.
The \Dz decay products are required to have $\ptot > 3\gevc$ and $\pt > 350\mevc$.
The \Dz decay products are required to form a vertex with a \chisq per degree of freedom (\chisqndf)  less than 10 and
a maximum distance of closest approach between any pair of \Dz decay products 
less than 0.12\mm. The \rich system is used to distinguish between kaons and 
pions when reconstructing the \Dz candidate. The \Dstarp candidates are reconstructed from 
\Dz candidates combined with a track with $\pt > 120\mevc$.
Decays are selected with candidate  
\Dz mass, \mD, of $1804 < \mD < 1924\mevcc$, where the notation \mD denotes the 
invariant mass of any of the considered final states; specific notations are used where appropriate. 
The difference, \deltam, in the reconstructed \Dstarp mass and \mD for candidate decays 
is required to be $137.9 < \deltam < 155.0\mevcc$.
The decay vertex of the \Dstar is constrained to coincide with the primary vertex~\cite{Hulsbergen:2005pu}. 

Differences in \Dstarp and \Dstarm meson production and detection efficiencies can 
introduce asymmetries across the phase-space distributions of the \Dz decay. 
To ensure that the soft pion is detected in the central region of the detector, fiducial cuts on its momentum are applied, as in Ref.~\cite{LHCb-PAPER-2011-023}.
The \Dz and \Dzb candidates are weighted by removing events so that they have same transverse momentum and pseudorapidity distributions.
To further cancel detection asymmetries the data set is selected to contain equal quantities of data collected with each magnetic field polarity. 
Events are randomly removed from the largest subsample of the two magnetic field polarity configurations.  

Each data sample is investigated for background contamination. The reconstructed \Dz mass is searched for evidence of backgrounds 
from misreconstructed \Dz decays in which \kaon/\pion misidentification has occurred. 
Candidates in which only a single final-state particle is misidentified are reconstructed outside the \mD signal range. No evidence for candidates with two, three, or four \kaon/\pion misidentifications is observed. Charm mesons from $b$-hadron decays are strongly suppressed by the requirement that the \Dz candidate originates from a primary vertex. This source of background is found to have a negligible contribution.

\section{Method}

\begin{figure}[htbp]
       \centering
     \subfloat{\label{fig:mass:KKPiPi:D}%
       \includegraphics[width=0.495\textwidth]{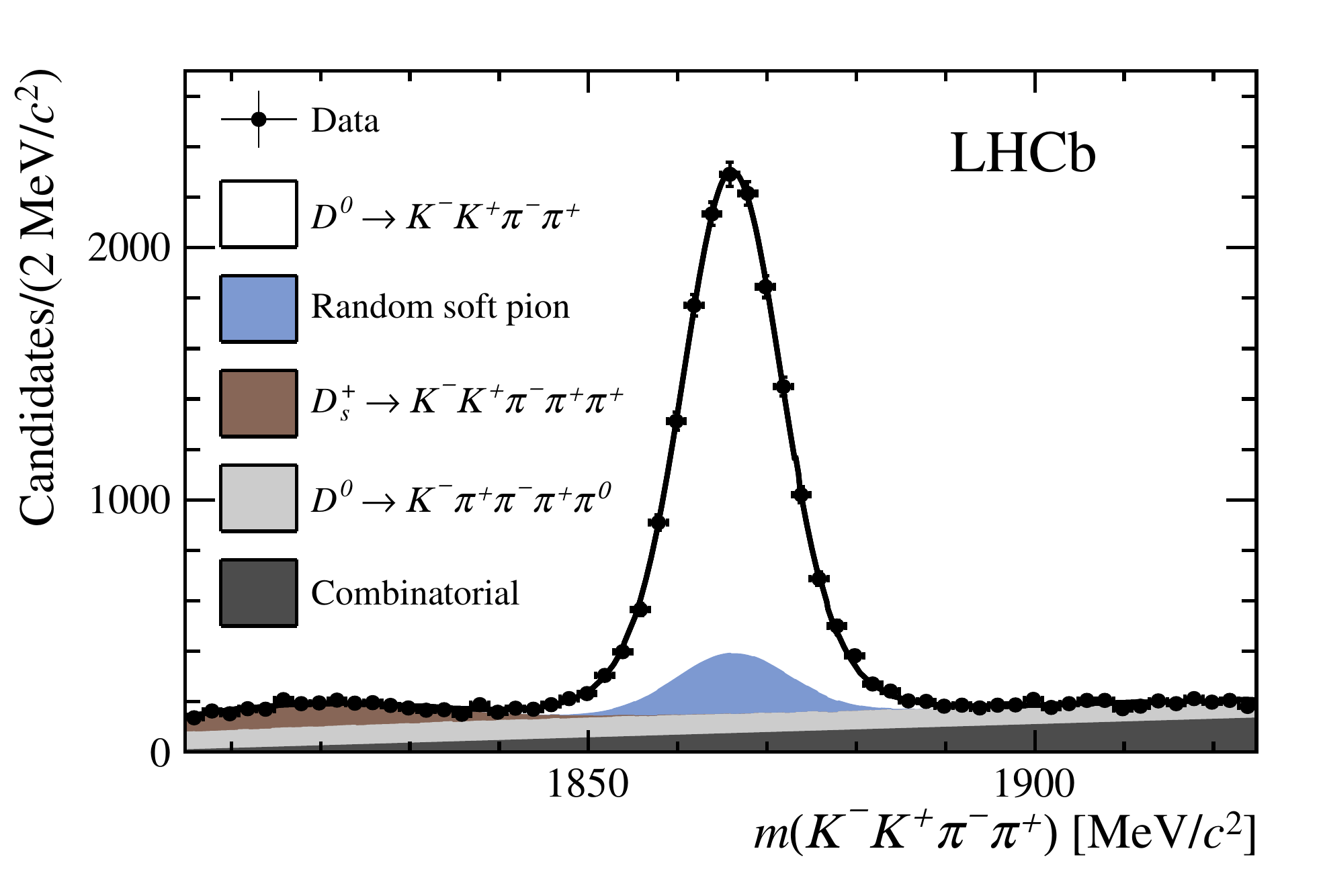}%
       \makebox[0cm][r]{\raisebox{0.205\textheight}[0cm]{\protect\subref{fig:mass:KKPiPi:D}}\hspace{0.04\textwidth}}
     }%
     \subfloat{\label{fig:mass:KKPiPi:DELTAM}%
       \includegraphics[width=0.495\textwidth]{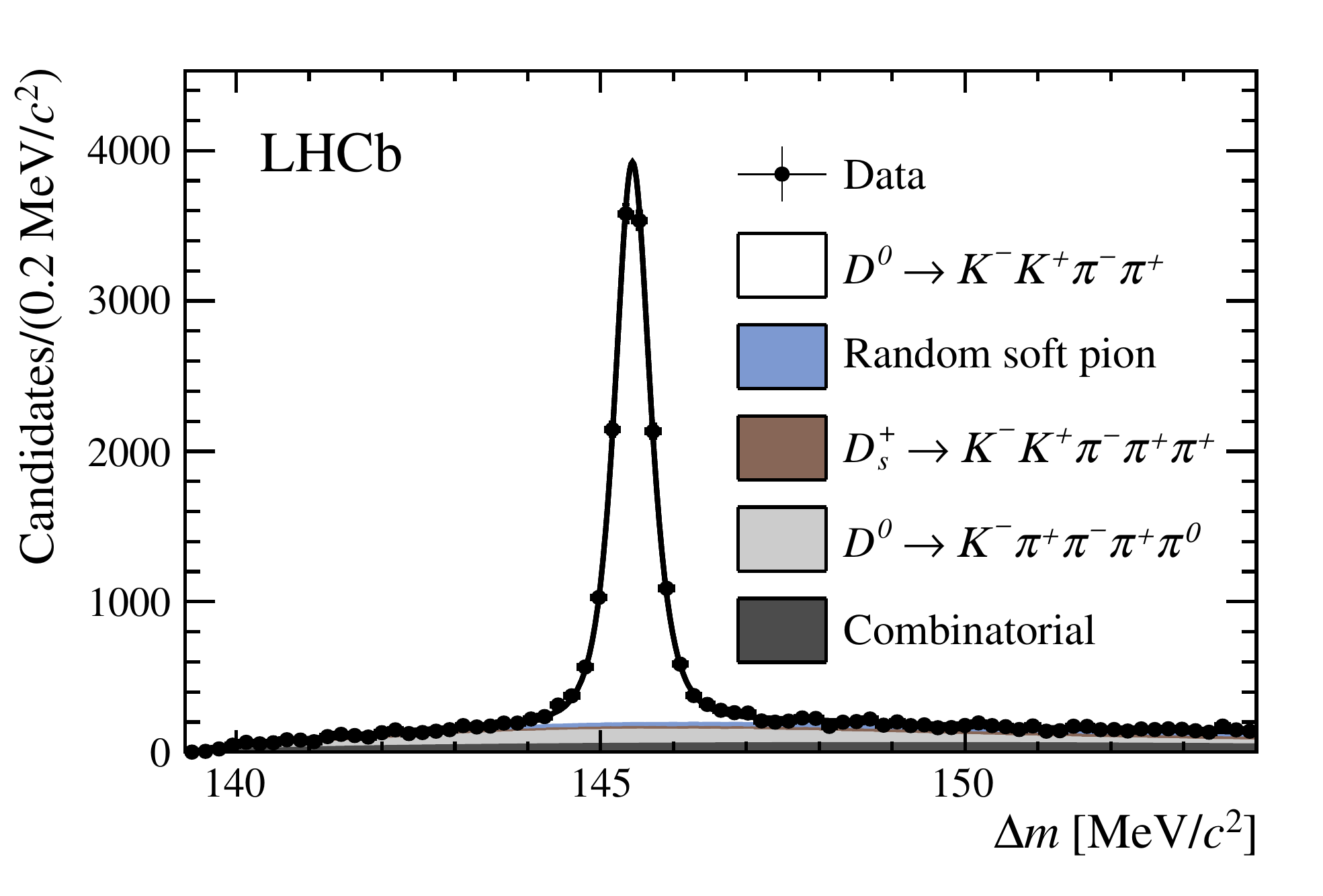}%
       \makebox[0cm][r]{\raisebox{0.205\textheight}[0cm]{\protect\subref{fig:mass:KKPiPi:DELTAM}}\hspace{0.04\textwidth}}
     }%

     \subfloat{\label{fig:mass:FourPi:D}%
       \includegraphics[width=0.495\textwidth]{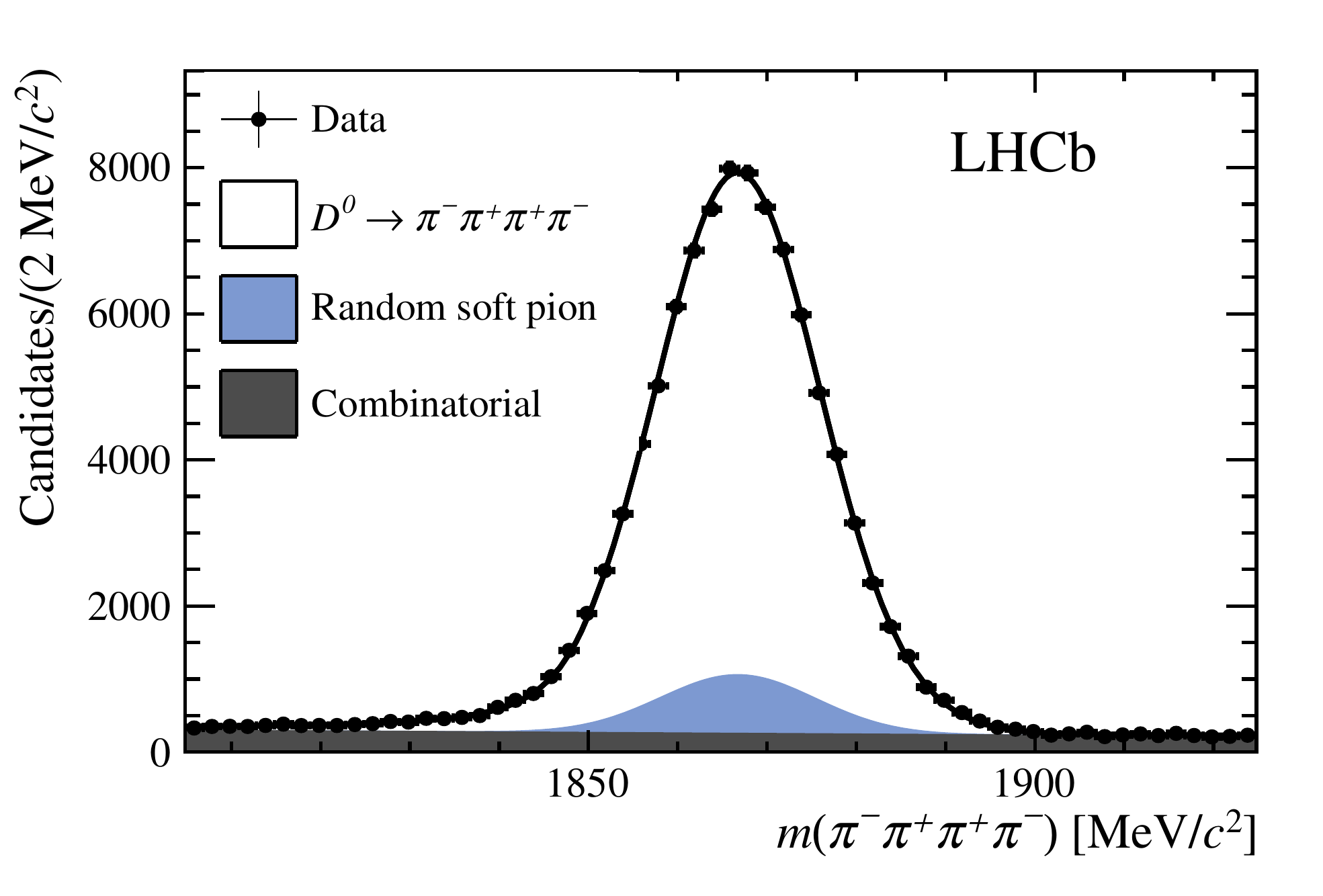}%
       \makebox[0cm][r]{\raisebox{0.205\textheight}[0cm]{\protect\subref{fig:mass:FourPi:D}}\hspace{0.04\textwidth}}
     }%
     \subfloat{\label{fig:mass:FourPi:DeltaM}%
       \includegraphics[width=0.495\textwidth]{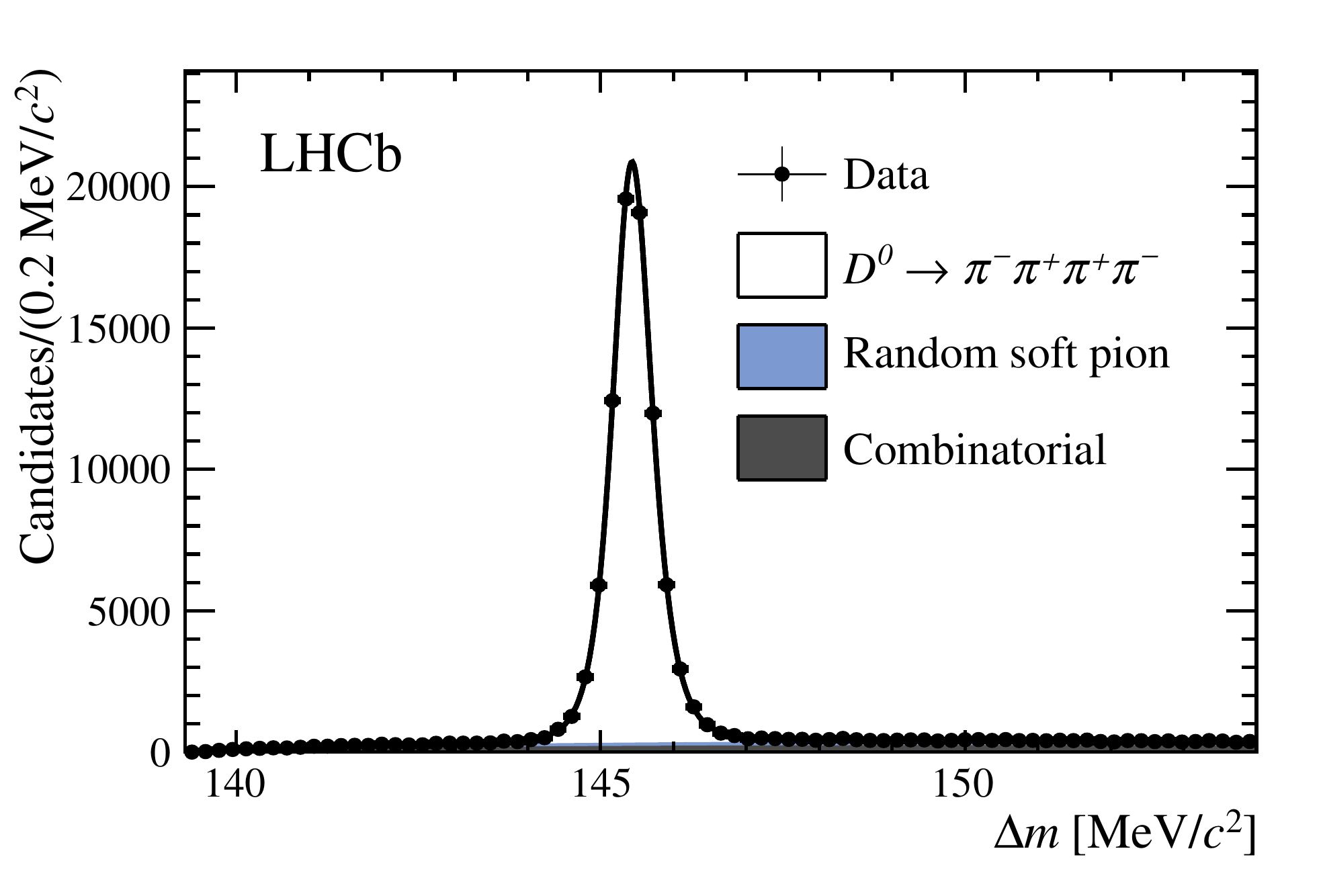}%
       \makebox[0cm][r]{\raisebox{0.205\textheight}[0cm]{\protect\subref{fig:mass:FourPi:DeltaM}}\hspace{0.04\textwidth}}
     }%

     \subfloat{\label{fig:mass:KThreePi:D}%
       \includegraphics[width=0.495\textwidth]{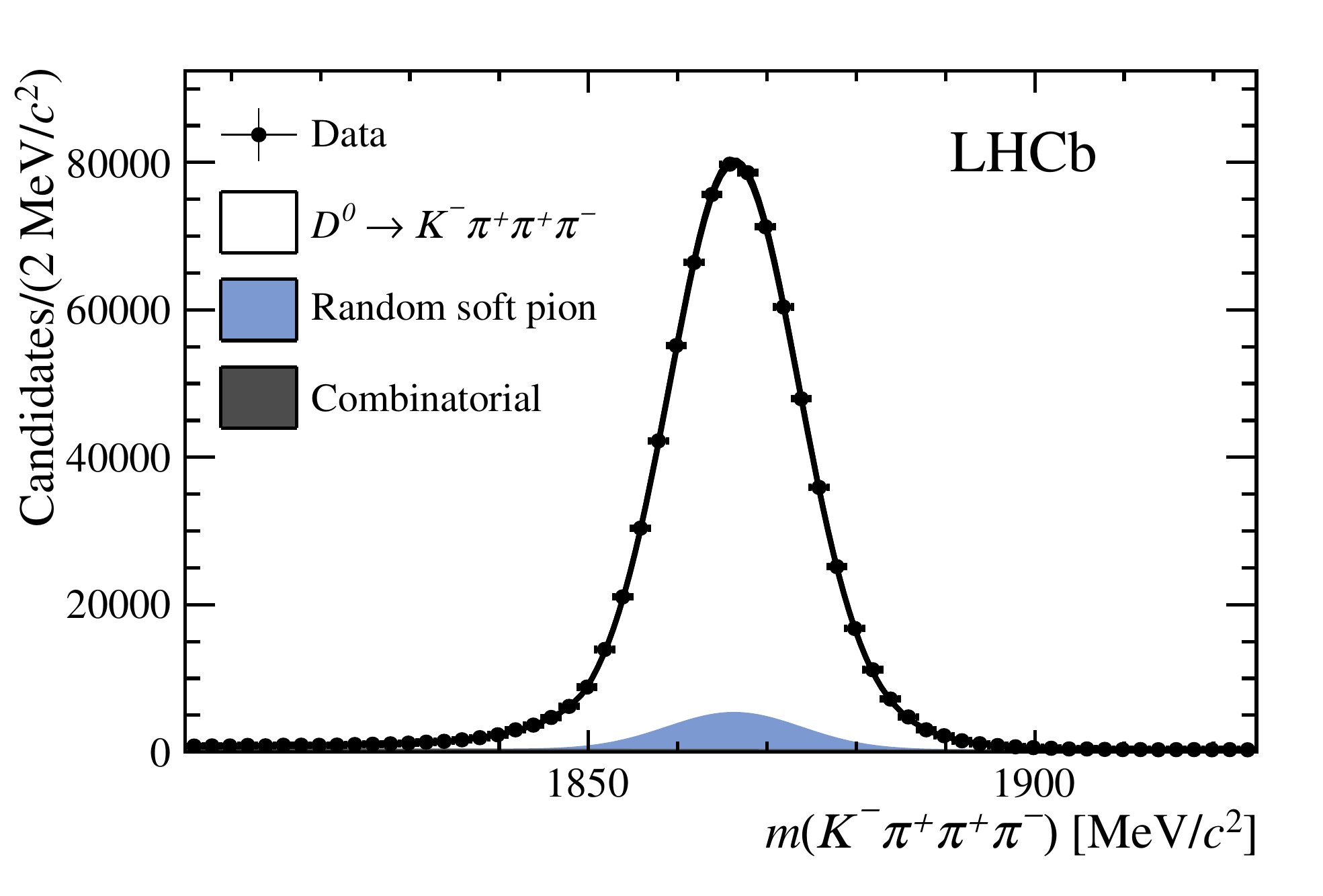}%
       \makebox[0cm][r]{\raisebox{0.205\textheight}[0cm]{\protect\subref{fig:mass:KThreePi:D}}\hspace{0.04\textwidth}}
     }%
     \subfloat{\label{fig:mass:KThreePi:DELTAM}%
       \includegraphics[width=0.495\textwidth]{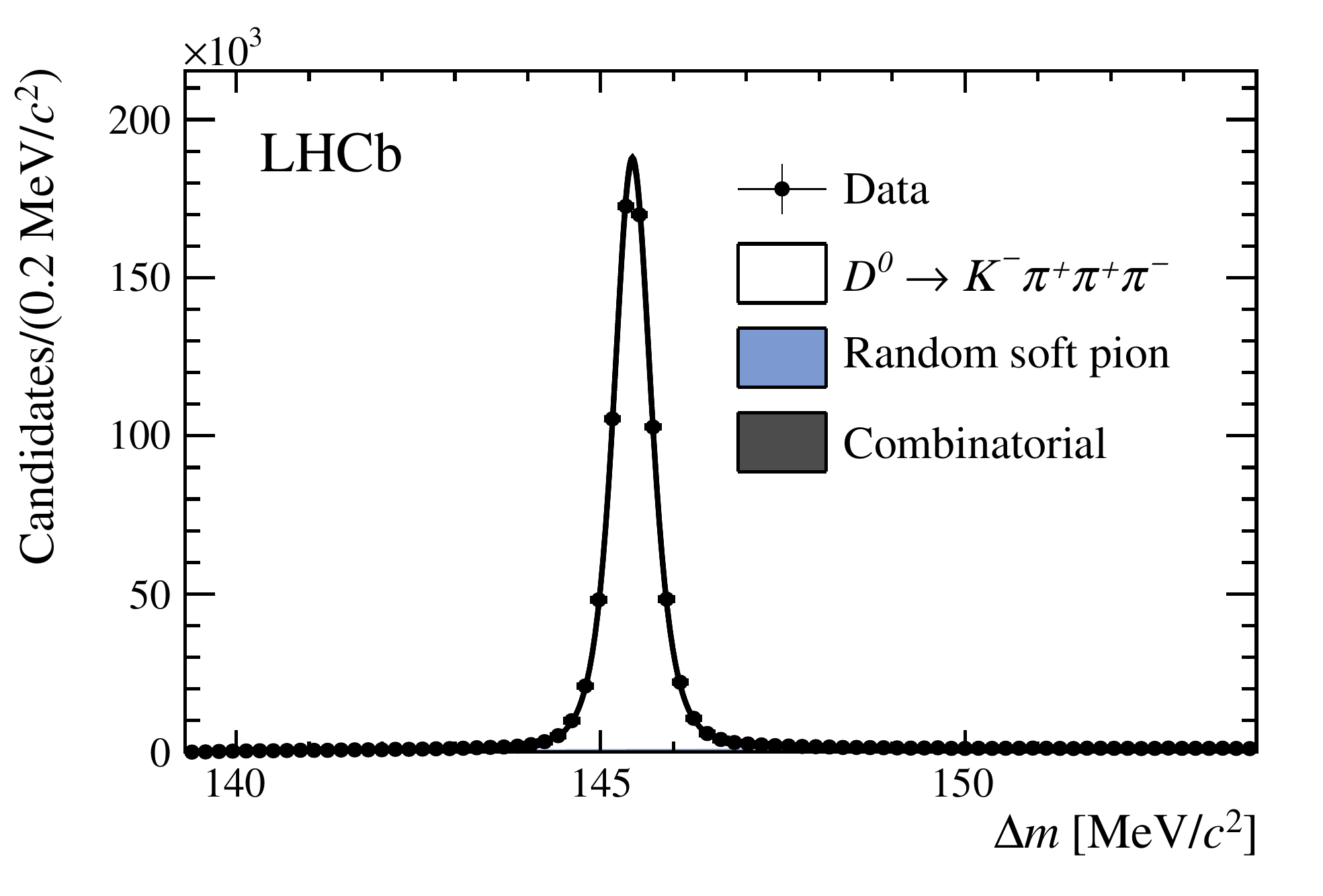}%
       \makebox[0cm][r]{\raisebox{0.205\textheight}[0cm]{\protect\subref{fig:mass:KThreePi:DELTAM}}\hspace{0.04\textwidth}}
     }%

      \caption{\small Distributions of (\protect\subref*{fig:mass:KKPiPi:D},\protect\subref*{fig:mass:FourPi:D},\protect\subref*{fig:mass:KThreePi:D}) \mD and (\protect\subref*{fig:mass:KKPiPi:DELTAM},\protect\subref*{fig:mass:FourPi:DeltaM},\protect\subref*{fig:mass:KThreePi:DELTAM}) \deltam  for (\protect\subref*{fig:mass:KKPiPi:D},\protect\subref*{fig:mass:KKPiPi:DELTAM}) \DKKPiPi, (\protect\subref*{fig:mass:FourPi:D},\protect\subref*{fig:mass:FourPi:DeltaM}) \DFourPi, and (\protect\subref*{fig:mass:KThreePi:D},\protect\subref*{fig:mass:KThreePi:DELTAM}) \DKThreePi candidates for magnet up polarity. Projections of the two-dimensional fits are overlaid, showing the contributions for signal, combinatorial background, and random soft pion background. The contributions from \DKPiPiPiPiPiZ and \DsKKPiPiPi contamination are also shown for the \DKKPiPi sample. \label{fig:method:massfits}}
\end{figure}
\afterpage{\clearpage}

\Fig{fig:method:massfits} shows the \mD and \deltam distributions for \Dz candidate decays to the final states \KKPiPi, \FourPi, and \KThreePi, for data taken with magnet up polarity. 
The distributions for \Dzb candidates and data taken with magnet down polarity are consistent with the distributions shown. 
Two-dimensional unbinned likelihood fits are made to the \mD and \deltam distributions to separate signal and background contributions. Each two-dimensional $[\mD, \Delta m]$ distribution includes contributions from the following sources: 
signal \Dz mesons from \Dstarp decays, which peak in both \mD and \deltam; combinatorial background candidates, which do not peak in 
either \mD or \deltam; background candidates from an incorrect association of a soft pion with a real \Dz meson, which peak in \mD and not in \deltam; incorrectly 
reconstructed \DsKKPiPiPi decays, which peak at low values of \mD but not in \deltam; and misreconstructed \DKPiPiPiPiPiZ decays, which have broad distributions in 
both \mD and \deltam. The signal distribution is described by a Johnson function~\cite{JohnsonFunction} in \deltam and a Crystal Ball function~\cite{Skwarnicki:1986xj} plus a Gaussian function, with a shared peak value, in \mD. The combinatorial background is modelled with a first-order polynomial in \mD, and the
background from \Dz candidates each associated with a random soft pion is modelled by a Gaussian distribution in \mD.
Both combinatorial and random soft pion backgrounds are modelled with a function of the form
\begin{equation}
\label{eq:backEq}
f(\deltam) = \left[  \left( \deltam - \deltam_{0}  \right)+ p_{1} \left(\deltam - \deltam_{0} \right)^{2} \right]^{a}
\end{equation}
in \deltam, where $\deltam_{0}$ is the kinematic threshold (fixed to the pion mass), and the parameters $p_{1}$ and $a$ are allowed to float. 

Partially reconstructed \DsKKPiPiPi decays, where a single pion is not reconstructed, are investigated with simulated decays. 
This background is modelled with a Gaussian distribution in \mD and with a function $f(\deltam)$ as defined in \eqn{eq:backEq}. Misreconstructed \DKPiPiPiPiPiZ decays where a single \kaon/\pion misidentification has occurred and where the \piz is not reconstructed are modelled 
with a shape from simulated decays. Other potential sources of background are found to be negligible.  

For each two-dimensional $[\mD, \deltam]$ distribution a fit is first performed to the background region, \mbox{$139 < \deltam < 143 \mevcc$} or \mbox{$ 149 < \deltam < 155 \mevcc$}, to obtain the shapes of the combinatorial and soft pion backgrounds. 
The \deltam components of these shapes are fixed and a two-dimensional fit is subsequently performed simultaneously over four samples (\Dz magnet up, \Dzb magnet up, \Dz magnet down, and \Dzb magnet down). The peak positions and widths of the signal shapes and all yields are allowed to vary independently for each sample, whilst all other parameters are shared among the four samples. Signal and background distributions are separated with the \sPlot \xspace statistical method~\cite{Pivk:2004ty}. 
The data sets contain $5.7 \times 10^{4}$ \DKKPiPi, $3.3 \times 10^{5}$ \DFourPi, and $2.9 \times 10^{6}$ \DKThreePi signal decays.

\begin{figure}[H]    \centering
  \subfloat{\label{fig:phase:KKPiPi:s12}%
   \includegraphics[width=0.333\textwidth]{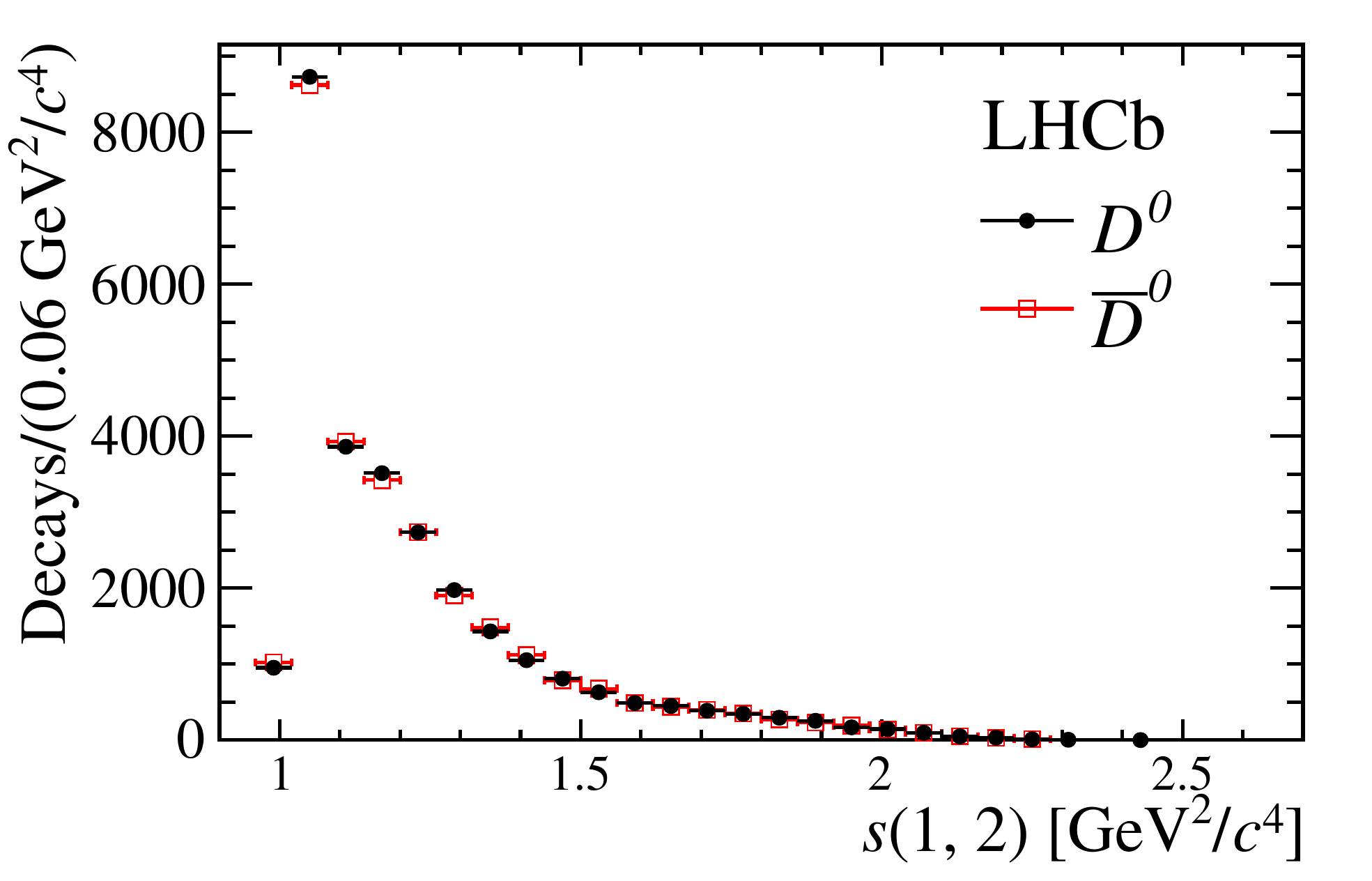} 
   }
  \subfloat{\label{fig:phase:KKPiPi:s23}%
    \includegraphics[width=0.333\textwidth]{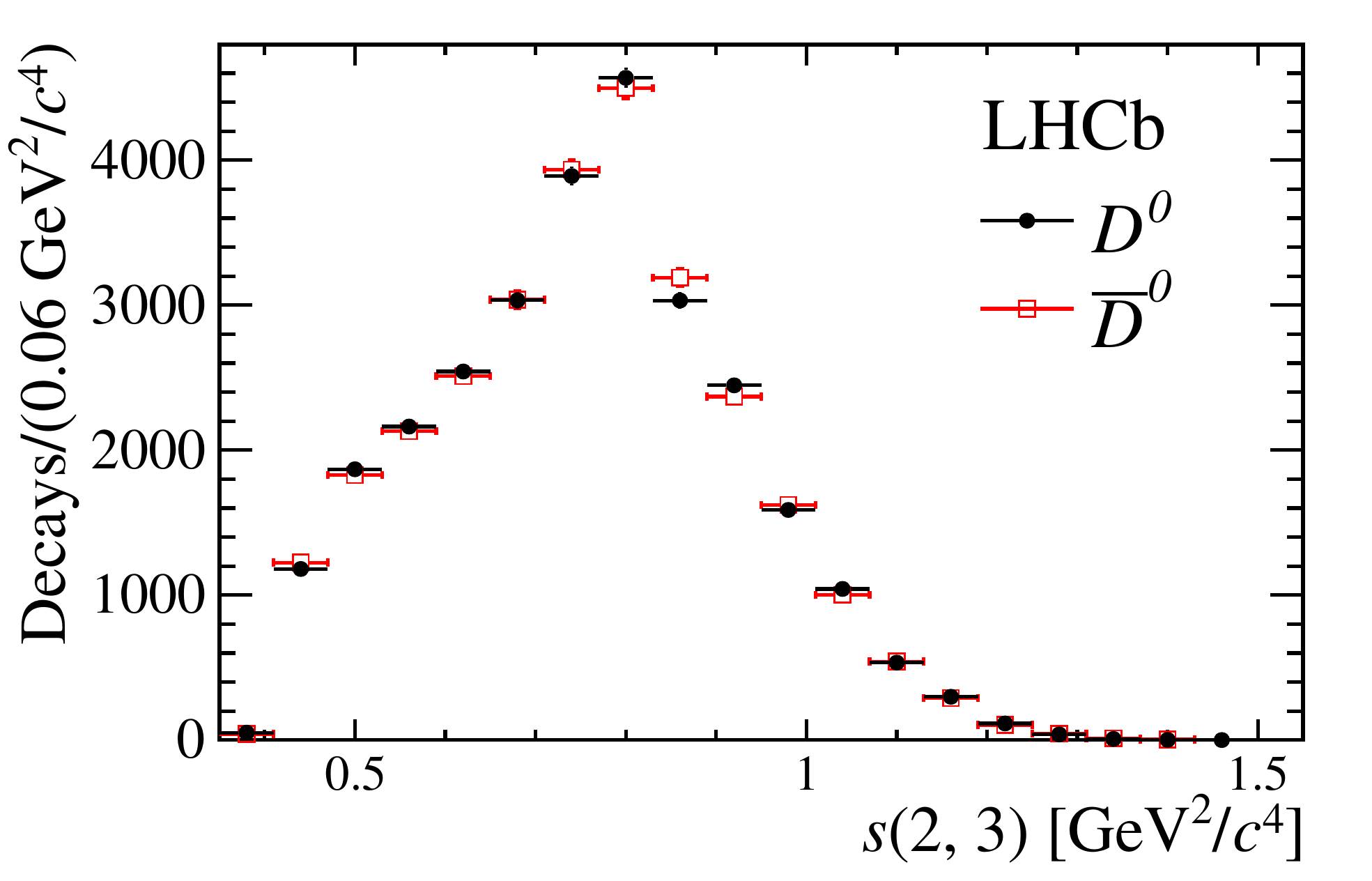} 
	}
  \subfloat{\label{fig:phase:KKPiPi:s123}%
   \includegraphics[width=0.333\textwidth]{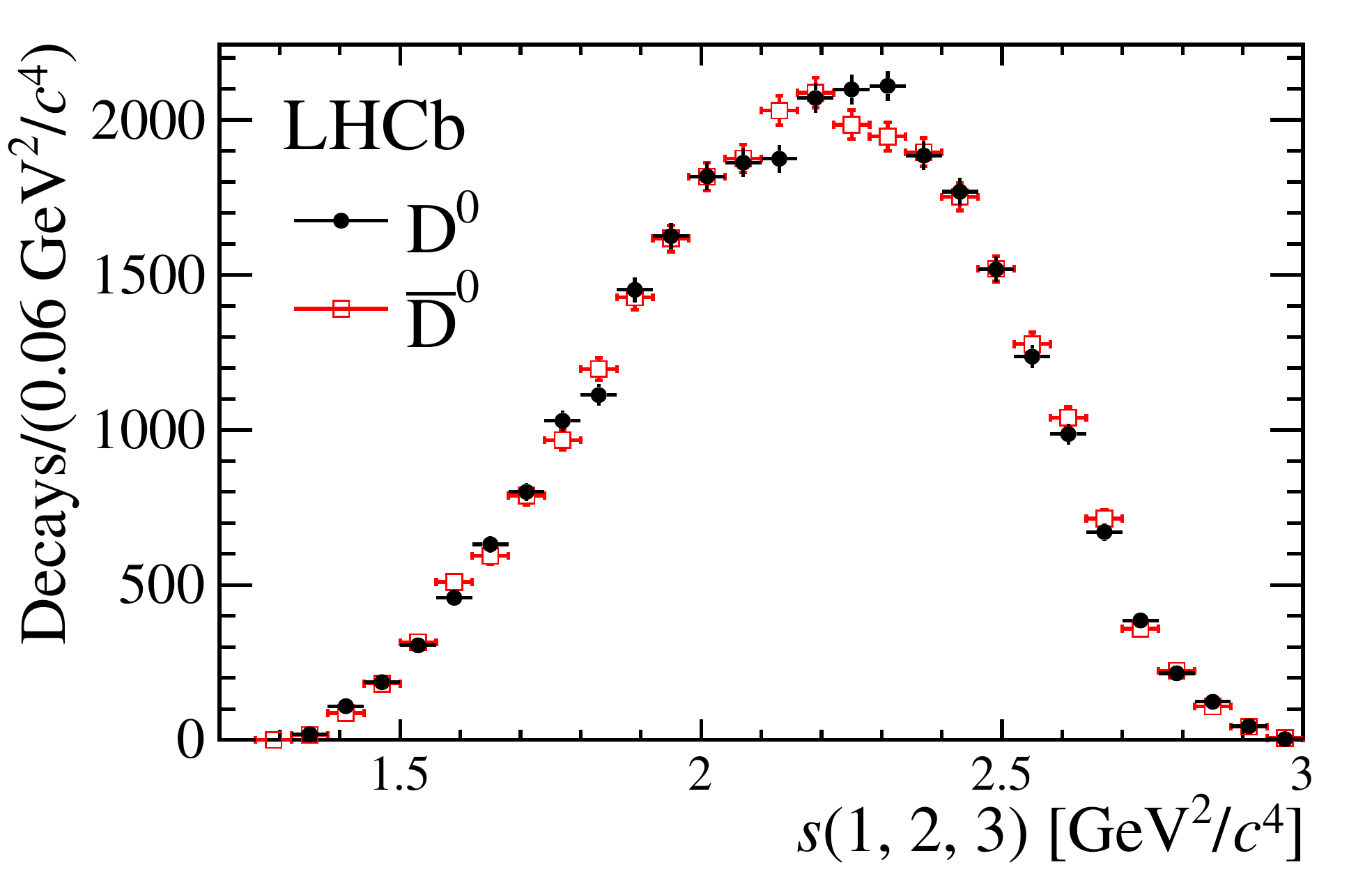} 
   }

  \subfloat{\label{fig:phase:KKPiPi:s234}%
   \includegraphics[width=0.333\textwidth]{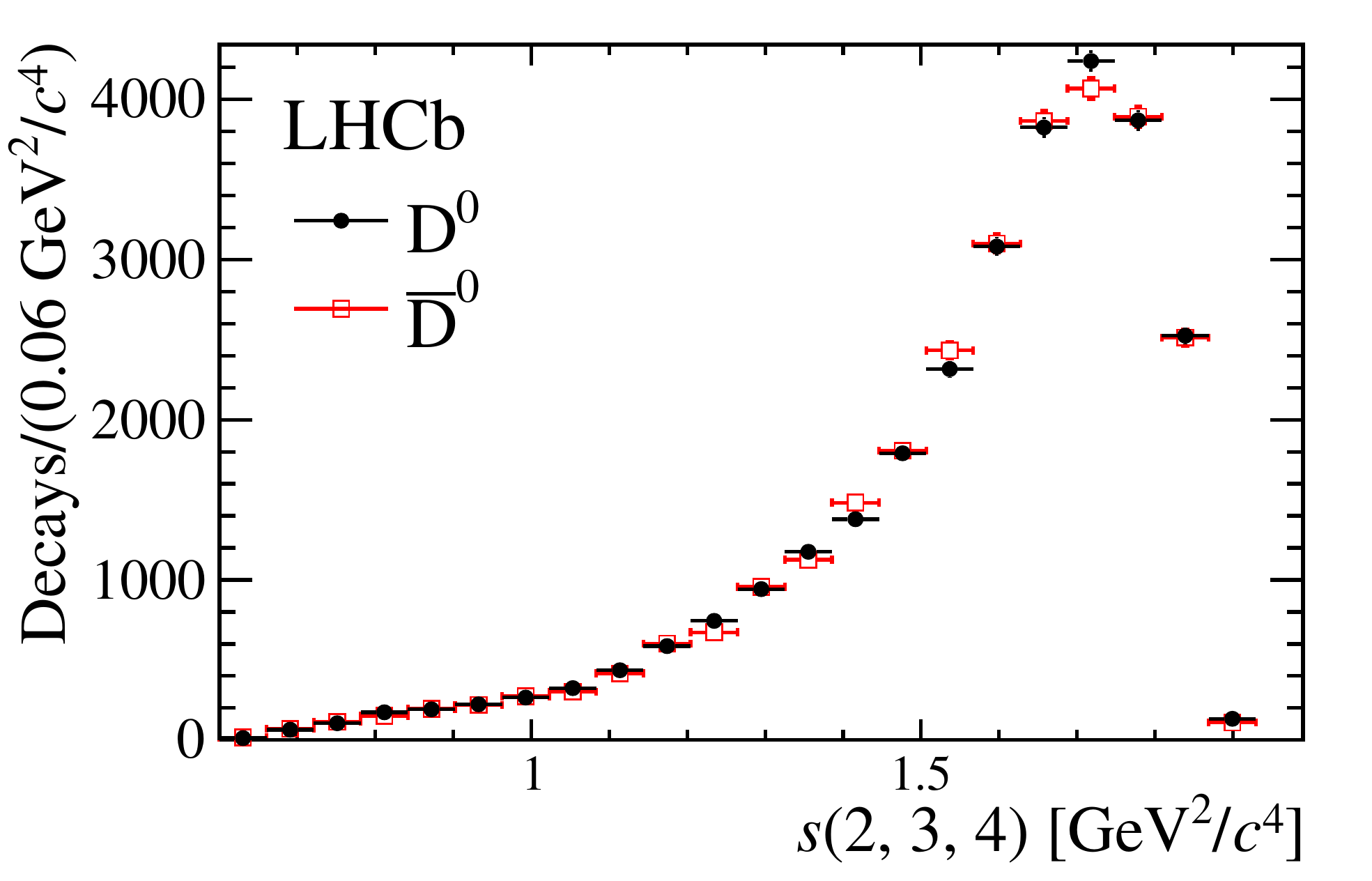} 
   }        
  \subfloat{\label{fig:phase:KKPiPi:s34}%
   \includegraphics[width=0.333\textwidth]{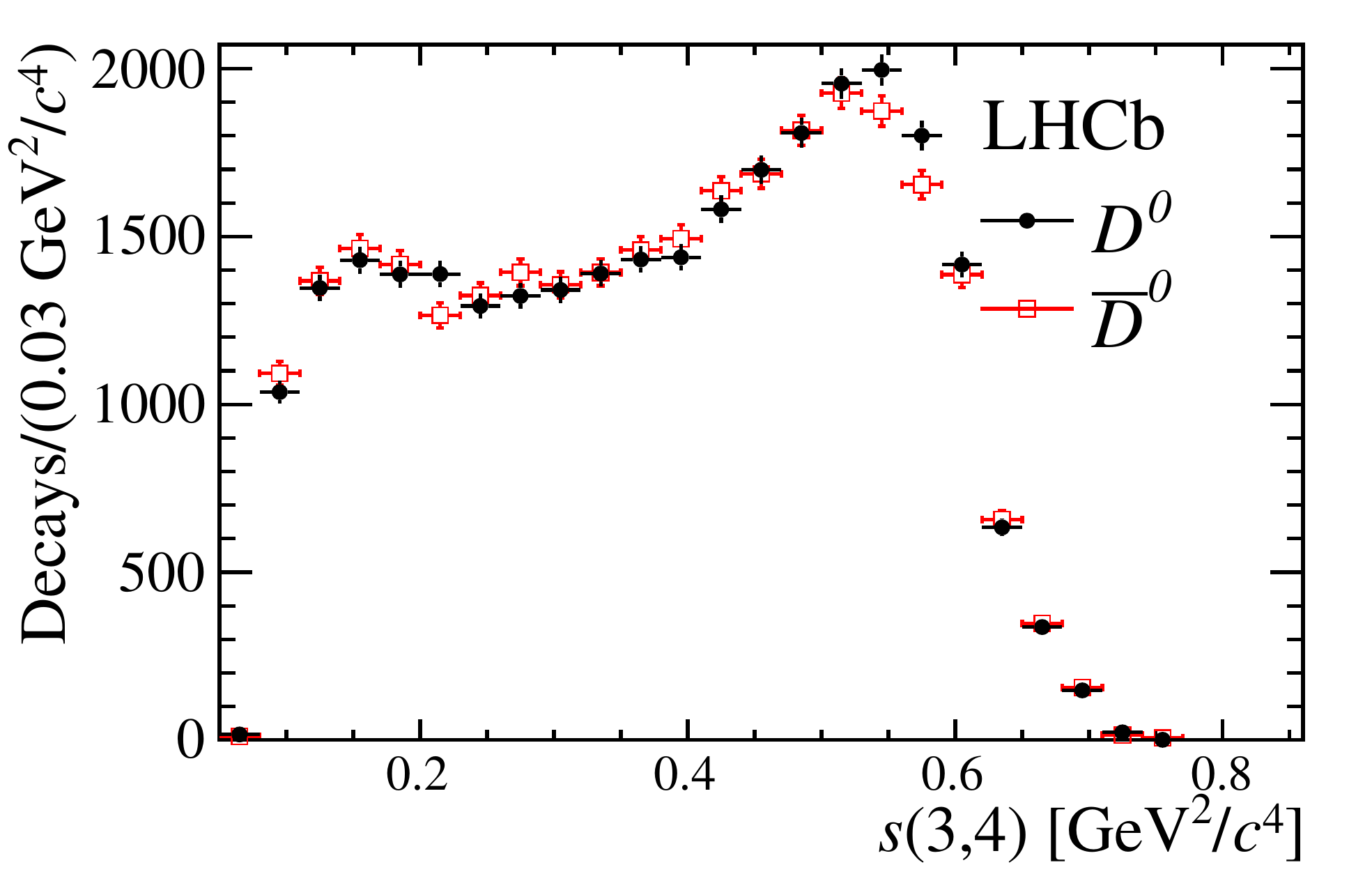} 
   }

   \caption{\small Invariant mass-squared distributions for \Dz meson (black, closed circles) and \Dzb meson (red, open squares) decays to the 
final state \KKPiPi. The invariant mass-squared combinations \invariantMassCombs correspond to \invariantMassCombsKKPiPi, respectively for the \Dz mode. 
The charge conjugate is taken for the \Dzb mode. 
The phase-space distribution of the \DKKPiPi decay is expected to be dominated by the quasi-two-body decay $\decay{\Dz}{\phi\rho^0}$ with additional contributions from  $\decay{\Dz}{\kaon_{1}(1270)^{\pm}\kaon^{\mp}}$ and $\decay{\Dz}{\kaon^{*}(1410)^\pm\kaon^{\mp}}$ decays~\cite{Artuso:2012df}. \label{fig:KKPiPi_Projections} }
\end{figure}

\begin{figure}[H] 
   \centering
  \subfloat{\label{fig:phase:FourPi:s12}%
   \includegraphics[width=0.333\textwidth]{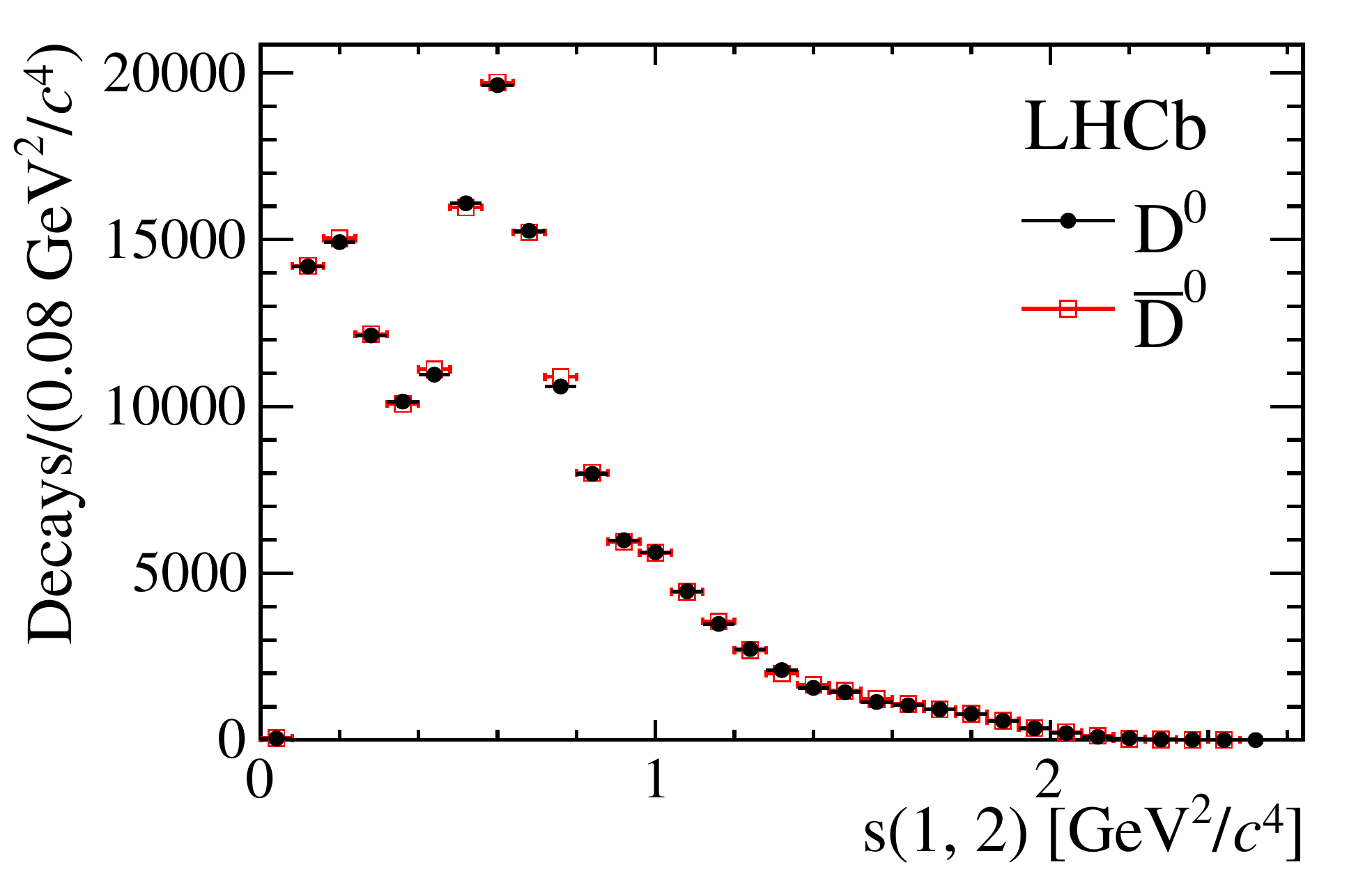} 
   }
  \subfloat{\label{fig:phase:FourPi:s23}%
    \includegraphics[width=0.333\textwidth]{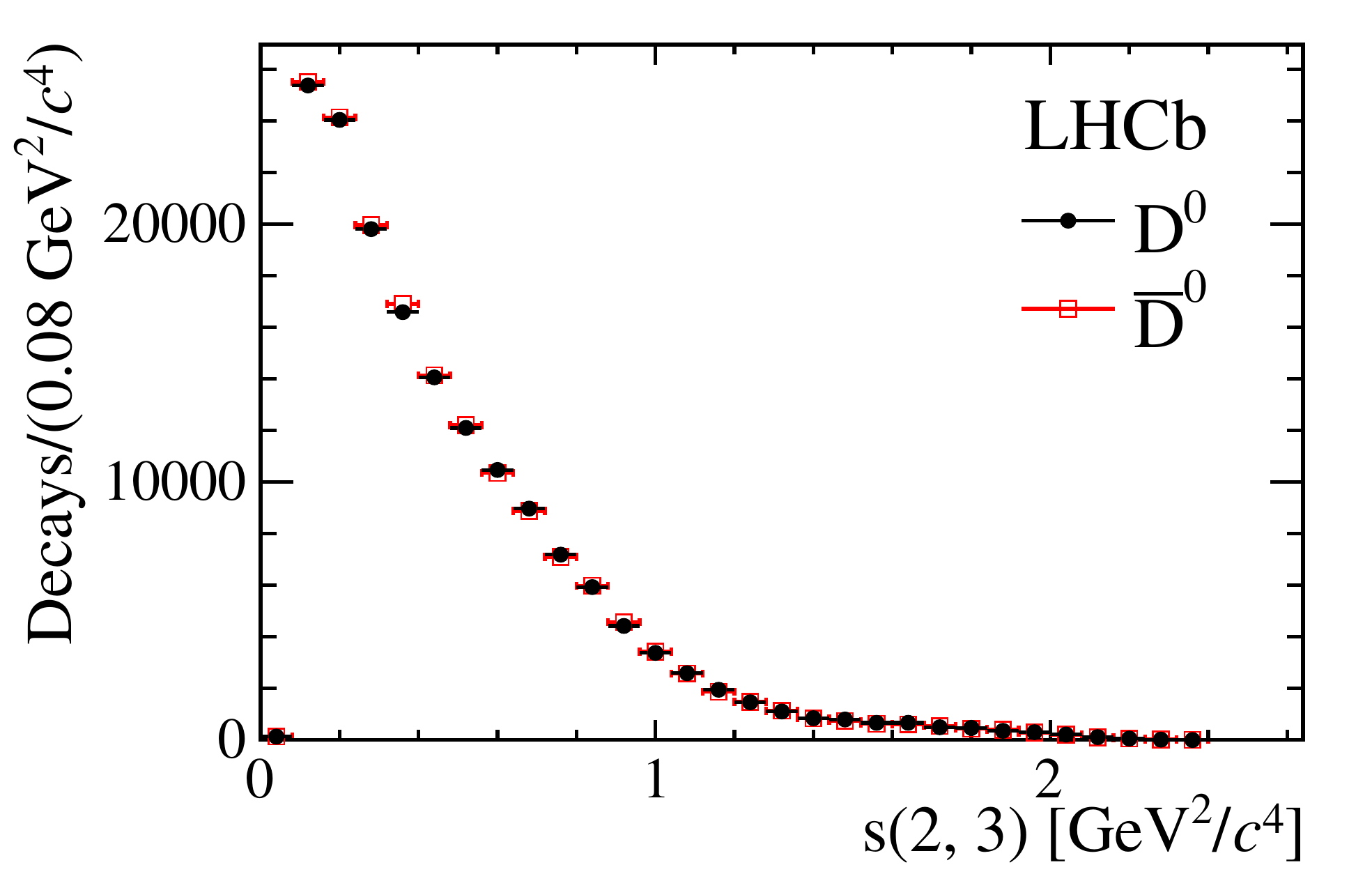} 
	}
  \subfloat{\label{fig:phase:FourPi:s123}%
   \includegraphics[width=0.333\textwidth]{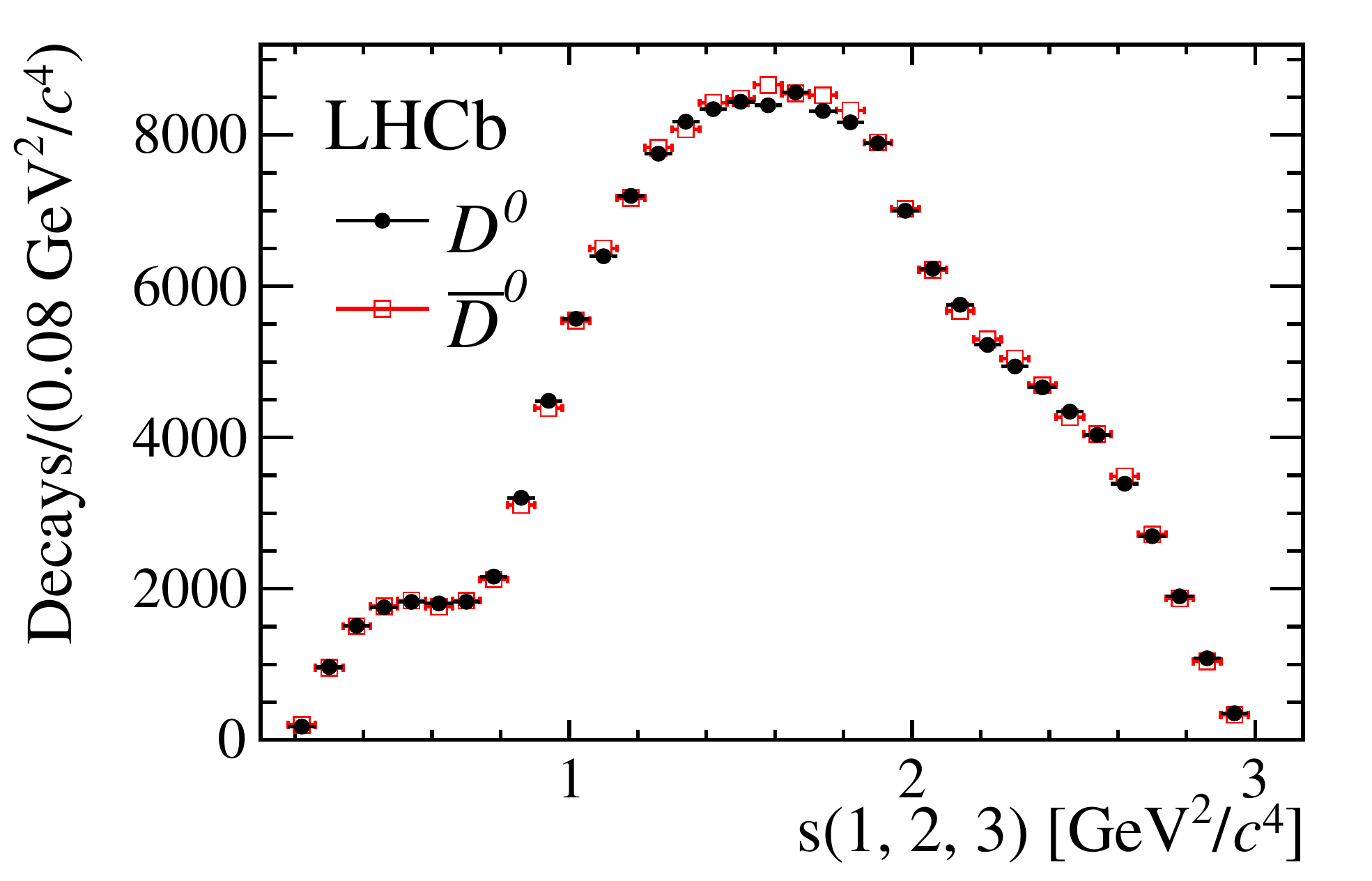} 
   }

   \caption{\small Invariant mass-squared distributions for \Dz meson (black, closed circles) and \Dzb meson (red, open squares) decays to the final state \FourPi. The invariant mass-squared combinations \invariantMassCombs correspond to \invariantMassCombsFourPi, respectively for the \Dz mode. 
The charge conjugate is taken for the \Dzb mode. Owing to the randomisation of the order of identical final-state particles the invariant mass-squared distributions s(2,3,4) and s(3,4) are statistically compatible with the invariant mass-squared distributions s(1,2,3) and s(1,2), respectively. As such the invariant mass-squared distributions s(2,3,4) and s(3,4) are not shown. The phase-space distribution of the \DFourPi decay is expected to be dominated by contributions from $\decay{\Dz}{a_{1}(1260)^{+} \pi^{-}}$ and $\decay{\Dz}{\rho^{0} \rho^{0}}$ decays~\cite{Link:2007fi}. 
 \label{fig:FourPi_Projections} }
\end{figure}
\afterpage{\clearpage}

The phase space of a spin-0 decay to four pseudoscalars can be described with five invariant mass-squared combinations: \invariantMassCombs, where the indices 1, 2, 3, and 4 correspond to the decay products of the \Dz meson following the ordering of the decay definitions. 
The ordering of identical final-state particles is randomised.

The rich amplitude structures are visible in the invariant mass-squared distributions for \Dz and \Dzb decays to the final states \KKPiPi and \FourPi, shown in \figs{fig:KKPiPi_Projections} and \ref{fig:FourPi_Projections}, respectively. The momenta of the final-state particles are calculated with the decay vertex of the \Dstar constrained to coincide with the primary vertex and the mass of the \Dz candidates constrained to the world average value of 1864.86\mevcc~\cite{PDG2012}. 

An adaptive binning algorithm is devised to partition the phase space of the decay into five-dimensional  hypercubes. 
The bins are defined such that each contains a similar number of candidates, resulting in fine bins around resonances and coarse bins across sparsely populated regions of phase space. 

For each phase-space bin, \SCPi, defined in \eqn{eq:SCP}, is calculated. The number of signal events in bin $i$, $N_{i}$, is calculated as the sum of the signal weights in bin $i$ and $\sigma_{i}^{2}$ is the sum of the squared weights. The normalisation factor, $\alpha$, is calculated as the ratio of the sum of the weights for \Dz candidates and the sum of the weights for  \Dzb candidates and is $1.001 \pm 0.008$, $0.996 \pm 0.003$, and $0.998 \pm 0.001$ for the final states \KKPiPi, \FourPi, and \KThreePi, respectively.

\section{Production and instrumental asymmetries}

Checks for remaining production or reconstruction asymmetries are 
carried out by comparing the phase-space distributions from a variety of data sets designed to test particle/antiparticle detection asymmetries and ``left/right'' detection asymmetries. The ``left'' direction is defined as the bending direction of a positively charged particle with the magnet up polarity. Asymmetries in the background are studied with weighted background 
candidates and mass sidebands. 

Left/right asymmetries in detection efficiencies are investigated by comparing the phase-space distributions of \Dz candidates in data 
taken with opposite magnet polarities, thus investigating the same flavour particles in opposite sides 
of the detector. 
Particle/antiparticle asymmetries are studied with the control channel \DKThreePi. The weighting based on \pt and pseudorapidity of the \Dz candidate and the normalisation across the phase space of the \Dz decay cancel the \Kp/\Km detection asymmetry in this control channel. The phase-space distribution of \Dz decays from data taken with one magnet polarity is compared with that of \Dzb decays from data taken with the opposite magnet polarity, for any sources of particle/antiparticle detection asymmetry, localised across the phase space of the \Dz decay.

The weighted distributions for each of the background components in the two-dimensional  fits are investigated for asymmetries in \DKKPiPi, \DFourPi, and \DKThreePi candidates. The \deltam and \mD  sidebands are also investigated to identify sources of asymmetry. 

The sensitivity to asymmetries is limited by the sample size, so \SCP is calculated only with statistical uncertainties.

\section{Sensitivity studies}

Pseudo-experiments are carried out to investigate the dependence of the sensitivity on the number of bins. 
Each pseudo-experiment is generated with a sample size comparable to that available in data.

\begin{figure}[p]
        \centering
	  \subfloat{\label{fig:Toy:SCP:NoCPV}%
		\includegraphics[width = 0.495\textwidth]{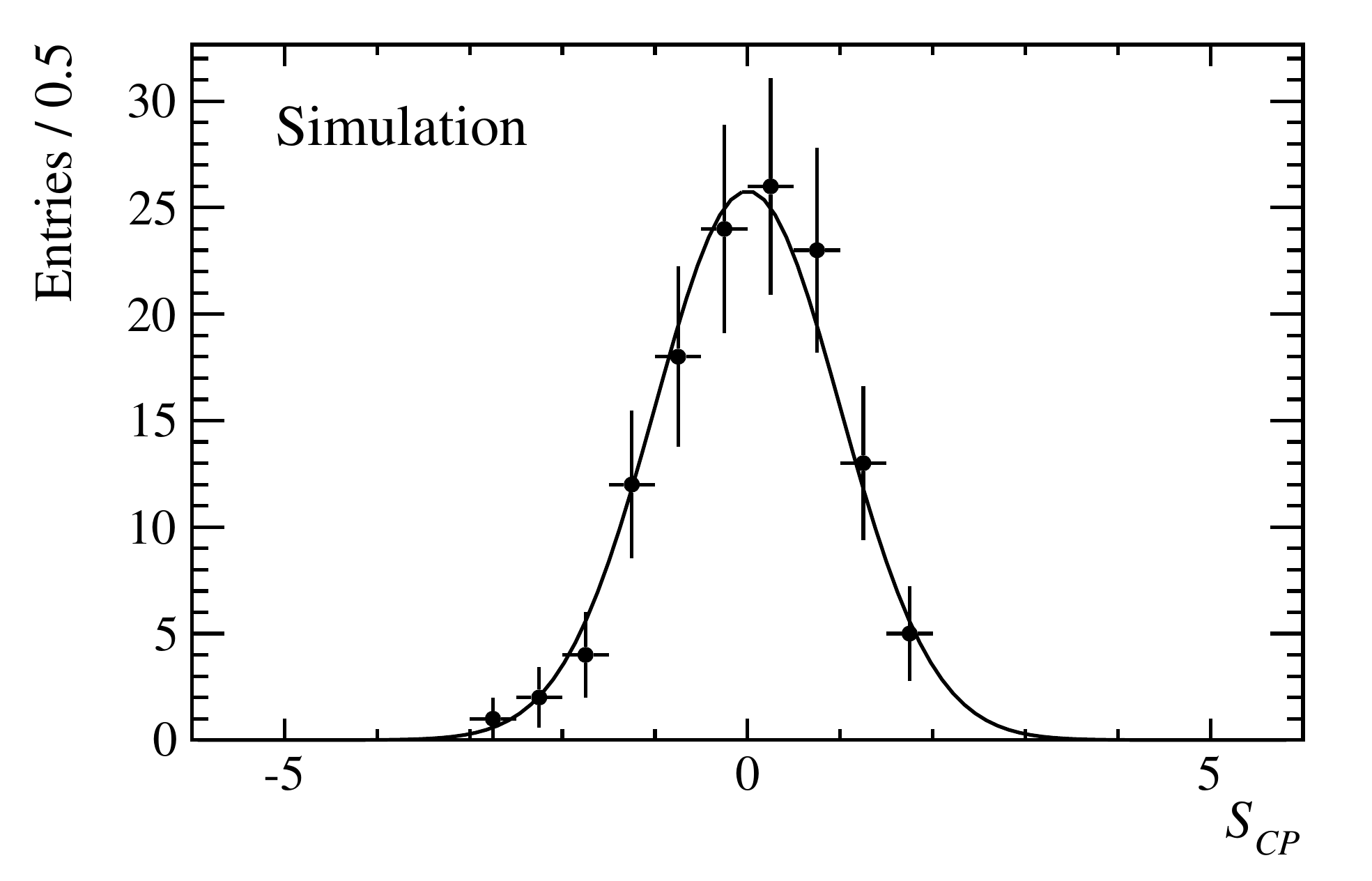}
		       \makebox[-0.2cm][r]{\raisebox{0.19\textheight}[0cm]{\protect\subref{fig:Toy:SCP:NoCPV}}\hspace{0.04\textwidth}}
		}
	  \subfloat{\label{fig:Toy:SCP:CPV}%
		\includegraphics[width = 0.495\textwidth]{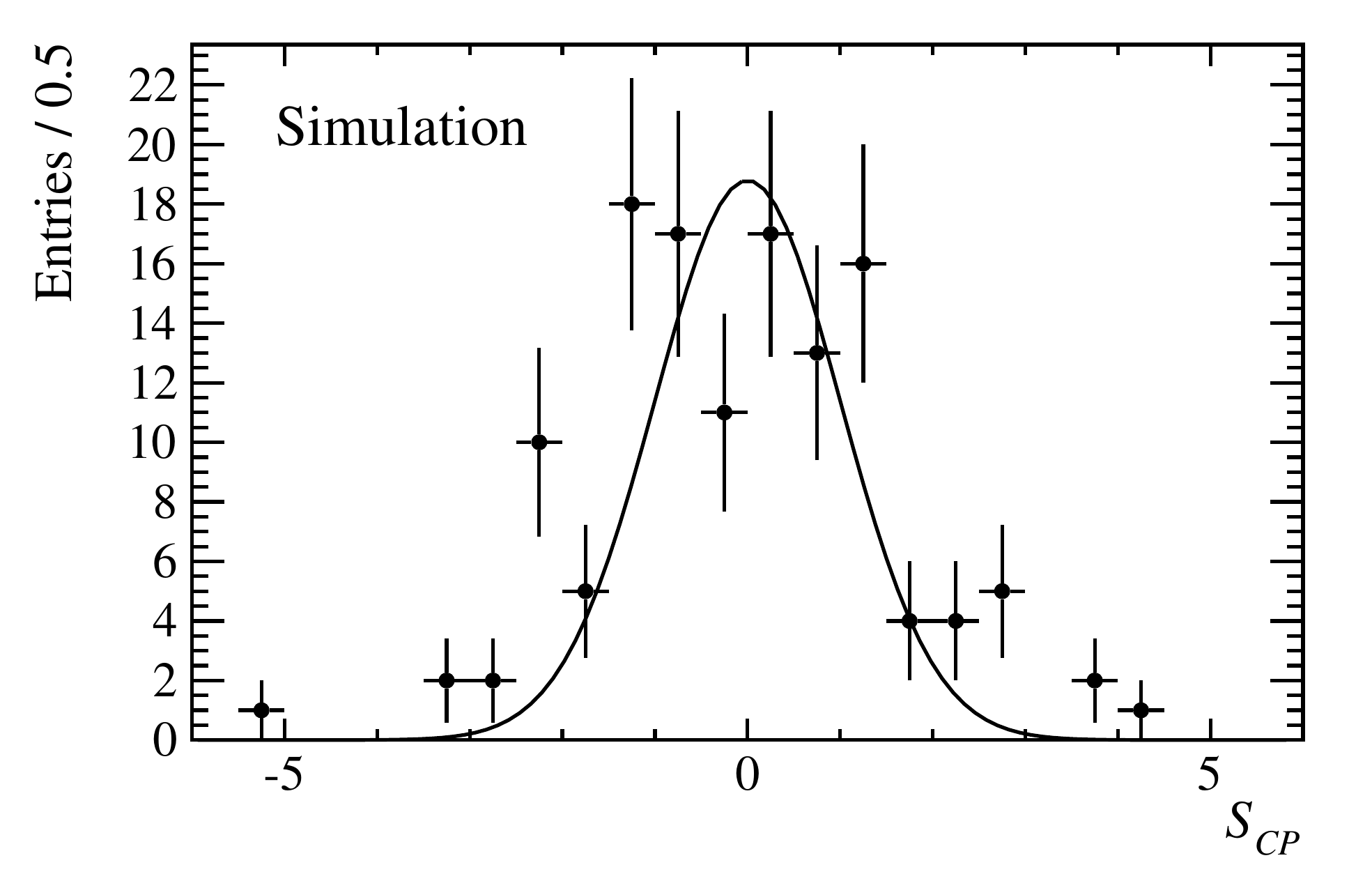}
		       \makebox[-0.2cm][r]{\raisebox{0.19\textheight}[0cm]{\protect\subref{fig:Toy:SCP:CPV}}\hspace{0.04\textwidth}}
	}
                \caption{\small{ Distributions of \SCP for \protect\subref{fig:Toy:SCP:NoCPV} a typical pseudo-experiment with generated \DFourPi decays without CPV  and for \protect\subref{fig:Toy:SCP:CPV} a typical pseudo-experiment with a  generated 10$^{\circ}$ phase difference between $\Dz \to a_{1}(1260)^{+} \pi^{-}$ and $\Dzb \to a_{1}(1260)^{-} \pi^{+}$ resonant decays. 
The points show the data distribution and the solid line is a reference Gaussian distribution corresponding to the no CPV hypothesis. 
The corresponding \pvalues under the hypothesis of no asymmetry for \protect\subref{fig:Toy:SCP:NoCPV} decays without CPV and \protect\subref{fig:Toy:SCP:CPV} decays with a 10$^{\circ}$ phase difference between $\Dz \to a_{1}(1260)^{+} \pi^{-}$ and $\Dzb \to a_{1}(1260)^{-} \pi^{+}$ resonant components are 85.6\% and $1.1\times10^{-16}$, respectively.
\label{fig:toy:SCP}}}
\end{figure}

Decays are generated with MINT, a software package for amplitude analysis of multi-body decays that has also been used by the \cleo collaboration~\cite{Artuso:2012df}. A sample of \DKKPiPi decays is generated according to the amplitude model reported by \cleo~\cite{Artuso:2012df}, and \DFourPi decays are generated according to the amplitude model from the FOCUS collaboration~\cite{Link:2007fi}. Phase and magnitude differences between \Dz and \Dzb decays are introduced. \Fig{fig:toy:SCP} shows the \SCP distributions for a typical pseudo-experiment  in which no CPV is present and for a typical pseudo-experiment with a phase difference of  $10^{\circ}$ between $\decay{\Dz}{a_{1}(1260)^{+} \pi^{-}}$ and $\decay{\Dzb}{a_{1}(1260)^{-} \pi^{+}}$ decays. 

Based on the results of the sensitivity study, a partition with 32 bins, with approximately 1800 signal events, is chosen for \DKKPiPi decays while a partition with 128 bins, 
with approximately 2500 signal events  is chosen for \DFourPi decays.
The \pvalues for the pseudo-experiments are uniformly distributed for the case of no CPV. 
The average \pvalue for a pseudo-experiment with a phase difference of 10$^{\circ}$ or a magnitude difference of 10$\%$ between 
$\decay{\Dz}{\phi \rho^0}$ and $\decay{\Dzb}{{\phi} {\rho}^0}$ decays for the \DKKPiPi mode and between
$\decay{\Dz}{a_{1}(1260)^{+} \pi^{-}}$ and $\decay{\Dzb}{a_{1}(1260)^{-} \pi^{+}}$ decays for the \DFourPi mode 
 is below $10^{-3}$. 

\section{Results}

\begin{table}[p]
\centering
\caption{ \small The \chisqndf and \pvalues under the hypothesis of no CPV for the control channel
\DKThreePi. The \pvalues are calculated separately for data samples taken with magnet up polarity, magnet down polarity, and the 
two polarities combined. \label{tab:PValueControl}}
\begin{tabular} {c  c   c  c}
& \pvalue (\%) (\chisqndf) & \pvalue (\%) (\chisqndf) & \pvalue (\%) (\chisqndf) \\
Bins & Magnet down & Magnet up & Combined sample \\ 
\hline
16 & 80.8 (10.2/15) & 21.2 (19.1/15) & 34.8 (16.5/15)\\
128  &  62.0 (121.5/127) &  75.9 (115.5/127) &  80.0 (113.4/127) \\
1024  &  27.5 (1049.6/1023) &  9.9 (1081.6/1023) &  22.1 (1057.5/1023) \\
\end{tabular}   
\end{table}
\afterpage{\clearpage}

\begin{table}[p]
\centering
\caption{\small The \chisqndf and \pvalues under the hypothesis of no CPV with three different partitions for
\DKKPiPi decays and \DFourPi decays. The \pvalues are calculated for a combined data sample with both data 
taken with magnet up polarity and data taken with magnet down polarity.  \label{tab:PValueResults} }
\begin{tabular}{c c c}
\multicolumn{3}{c}{\DKKPiPi}\\
Bins & \pvalue (\%) & \chisqndf\\
\hline
16  &  9.1 & 22.7/15 \\
32  &   9.1 & 42.0/31 \\
64  &  13.1  & 75.7/63 \\
\end{tabular}   
\quad
\quad
\quad
\begin{tabular}{c c c}\multicolumn{3}{c}{\DFourPi}\\
Bins & \pvalue (\%) & \chisqndf\\
\hline
64 & 28.8 & 68.8/63 \\
128  & 41.0  & 130.0/127 \\
256 & 61.7  & 247.7/255 \\
\end{tabular}   
 \end{table}
\afterpage{\clearpage}

\begin{figure}[p]
        \centering
     \subfloat{\label{fig:SCP:KKPiPi}%
       \includegraphics[width=0.495\textwidth]{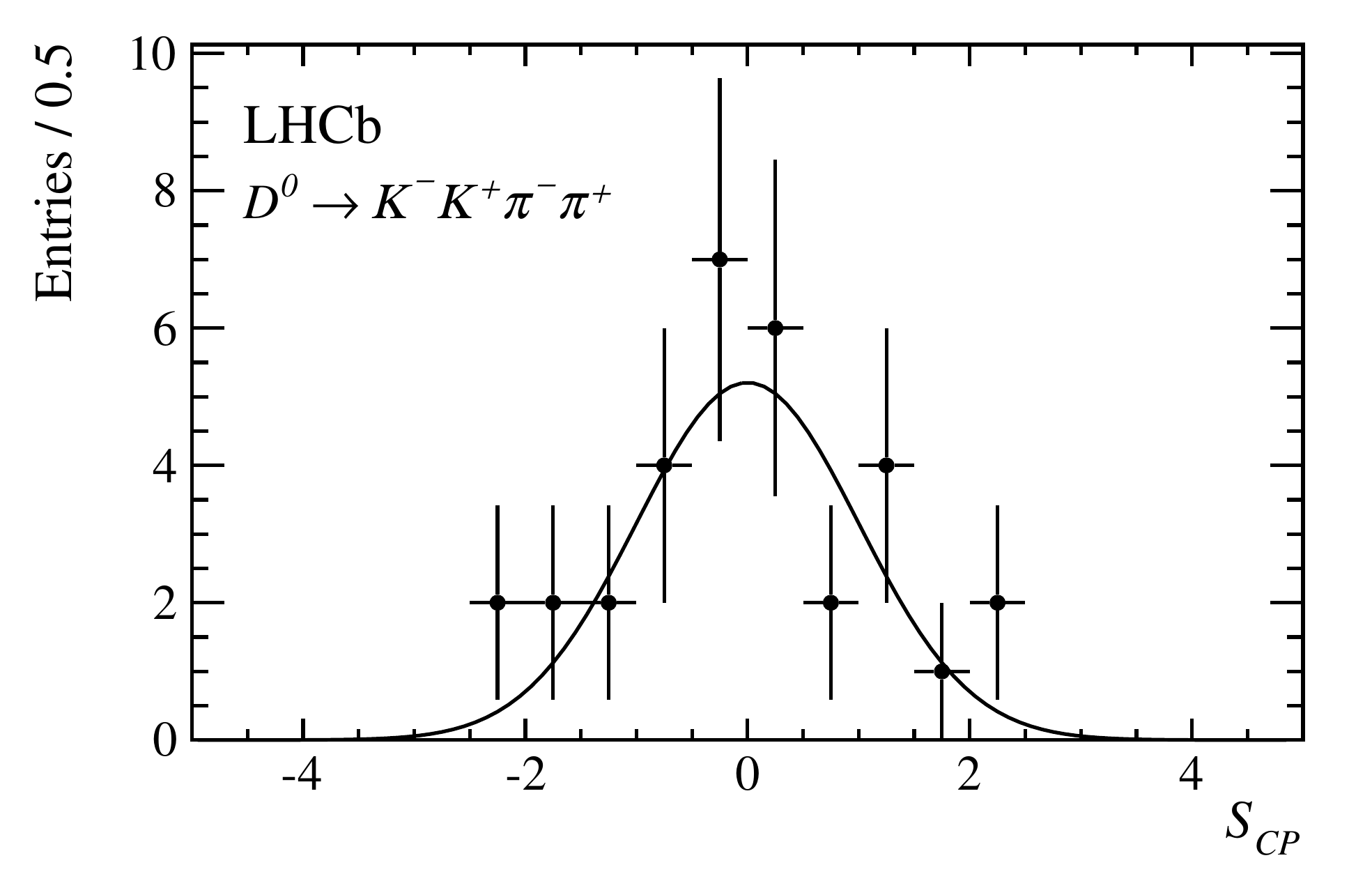}%
       \makebox[0cm][r]{\raisebox{0.19\textheight}[0cm]{\protect\subref{fig:SCP:KKPiPi}}\hspace{0.04\textwidth}}
     }%
     \subfloat{\label{fig:RAW:KKPiPi}%
       \includegraphics[width=0.495\textwidth]{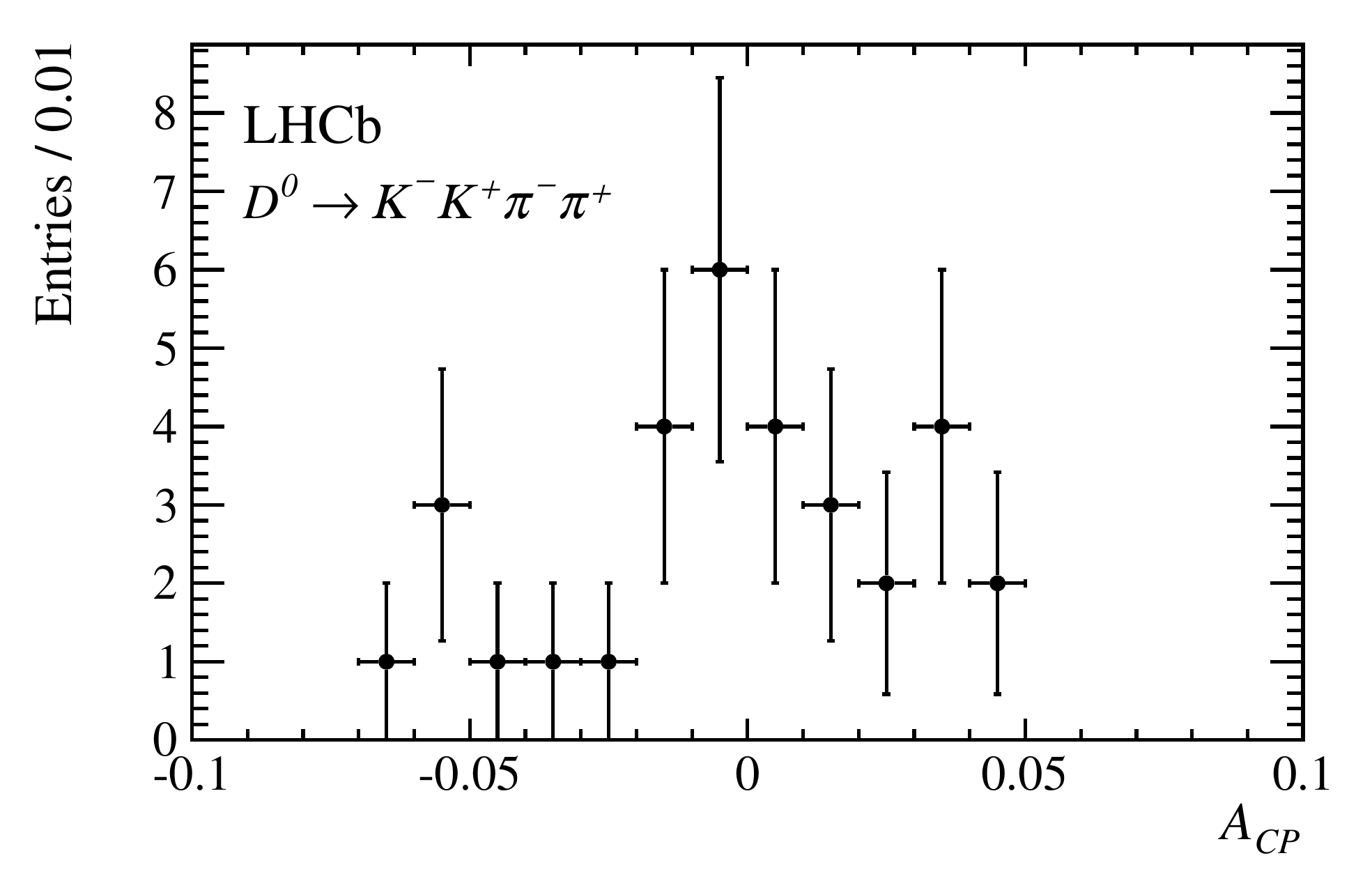}%
       \makebox[0cm][r]{\raisebox{0.19\textheight}[0cm]{\protect\subref{fig:RAW:KKPiPi}}\hspace{0.04\textwidth}}
     }%

     \subfloat{\label{fig:SCP:FourPi}%
       \includegraphics[width=0.495\textwidth]{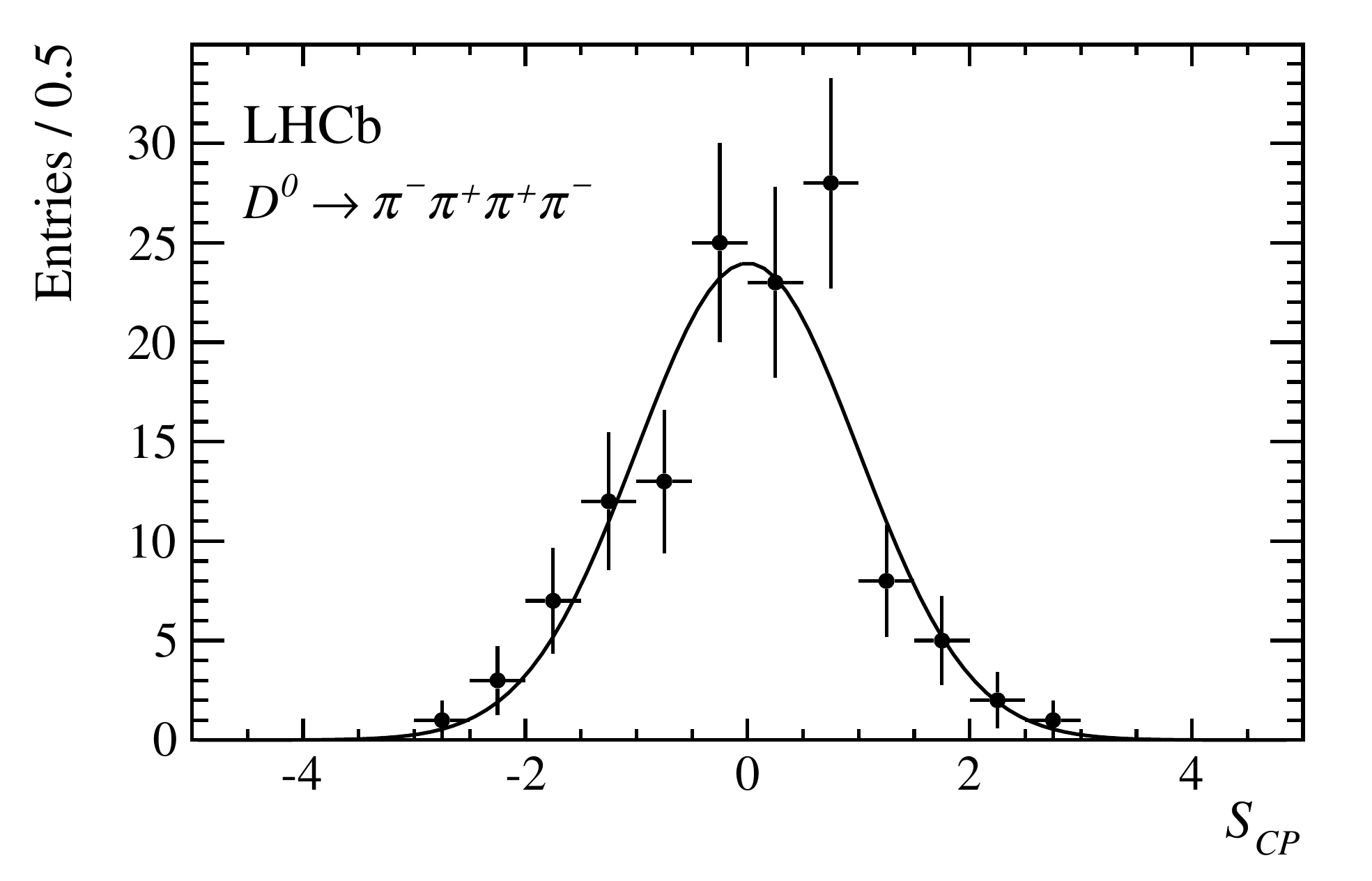}%
       \makebox[0cm][r]{\raisebox{0.19\textheight}[0cm]{\protect\subref{fig:SCP:FourPi}}\hspace{0.04\textwidth}}
     }%
     \subfloat{\label{fig:RAW:FourPi}%
       \includegraphics[width=0.495\textwidth]{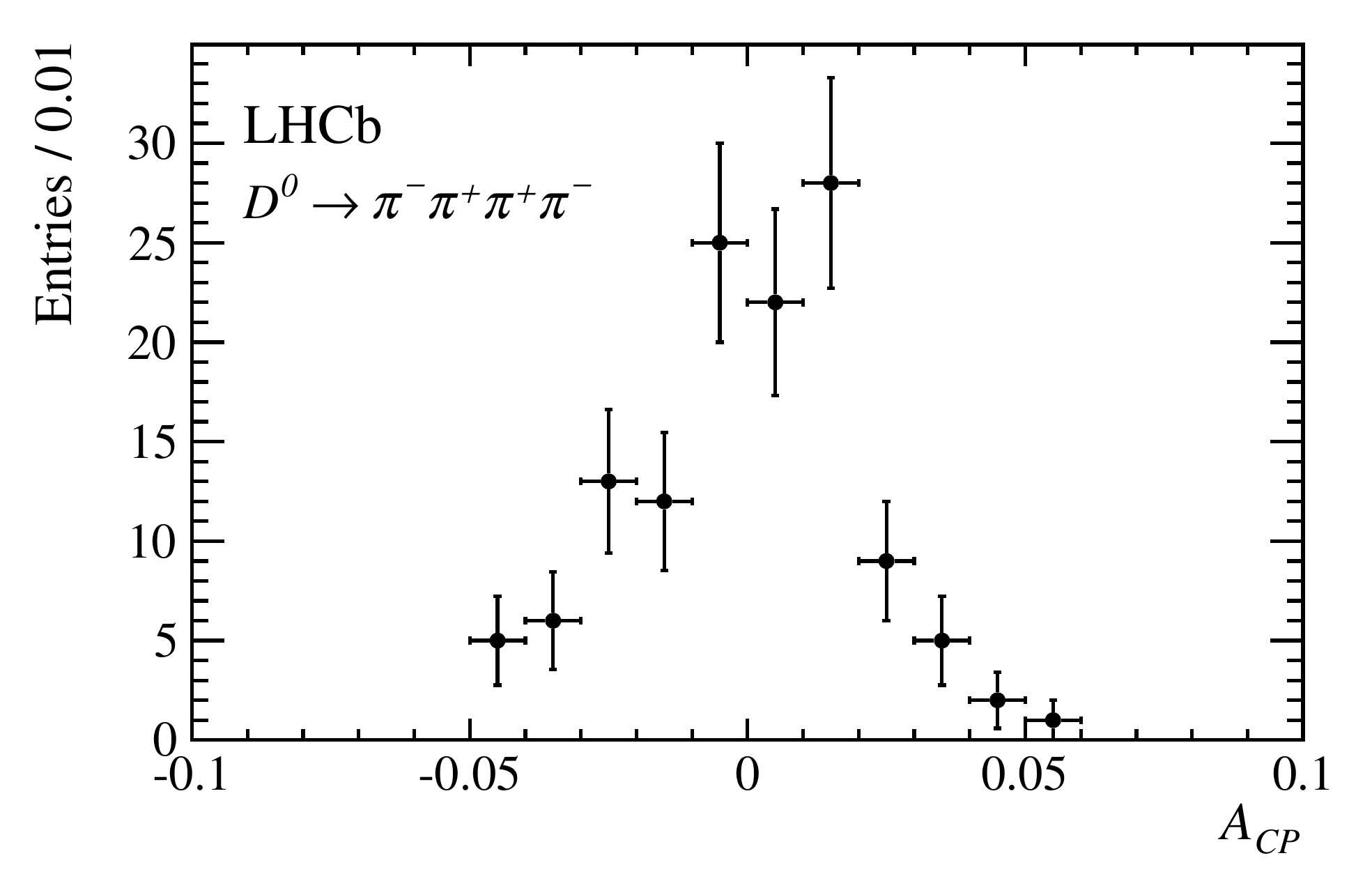}%
       \makebox[0cm][r]{\raisebox{0.19\textheight}[0cm]{\protect\subref{fig:RAW:FourPi}}\hspace{0.04\textwidth}}
     }%

  \subfloat{\label{fig:control:SCP}%
       \includegraphics[width=0.495\textwidth]{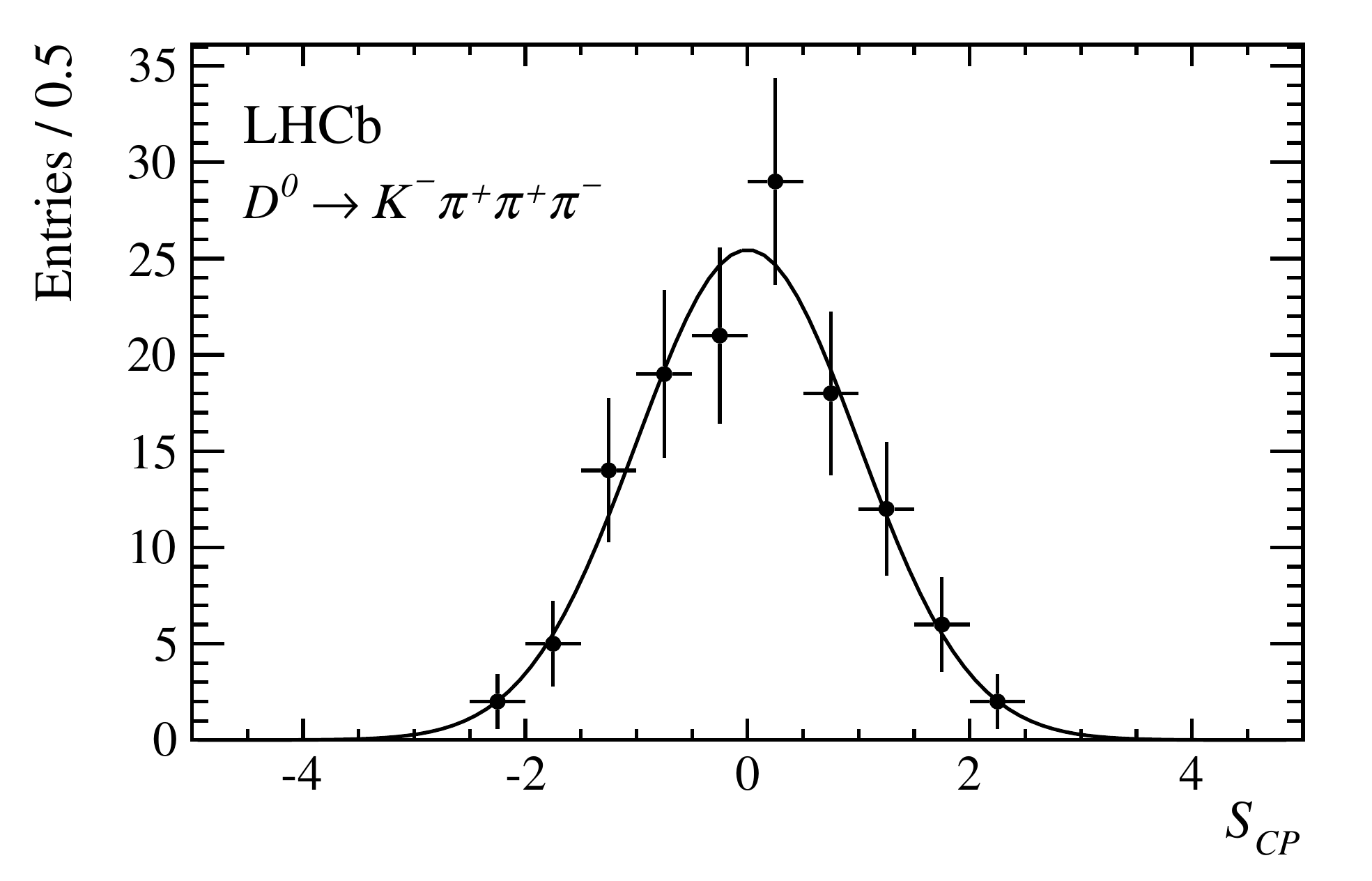}%
       \makebox[0cm][r]{\raisebox{0.19\textheight}[0cm]{\protect\subref{fig:control:SCP}}\hspace{0.04\textwidth}}
     }%
  \subfloat{\label{fig:control:RAW}%
       \includegraphics[width=0.495\textwidth]{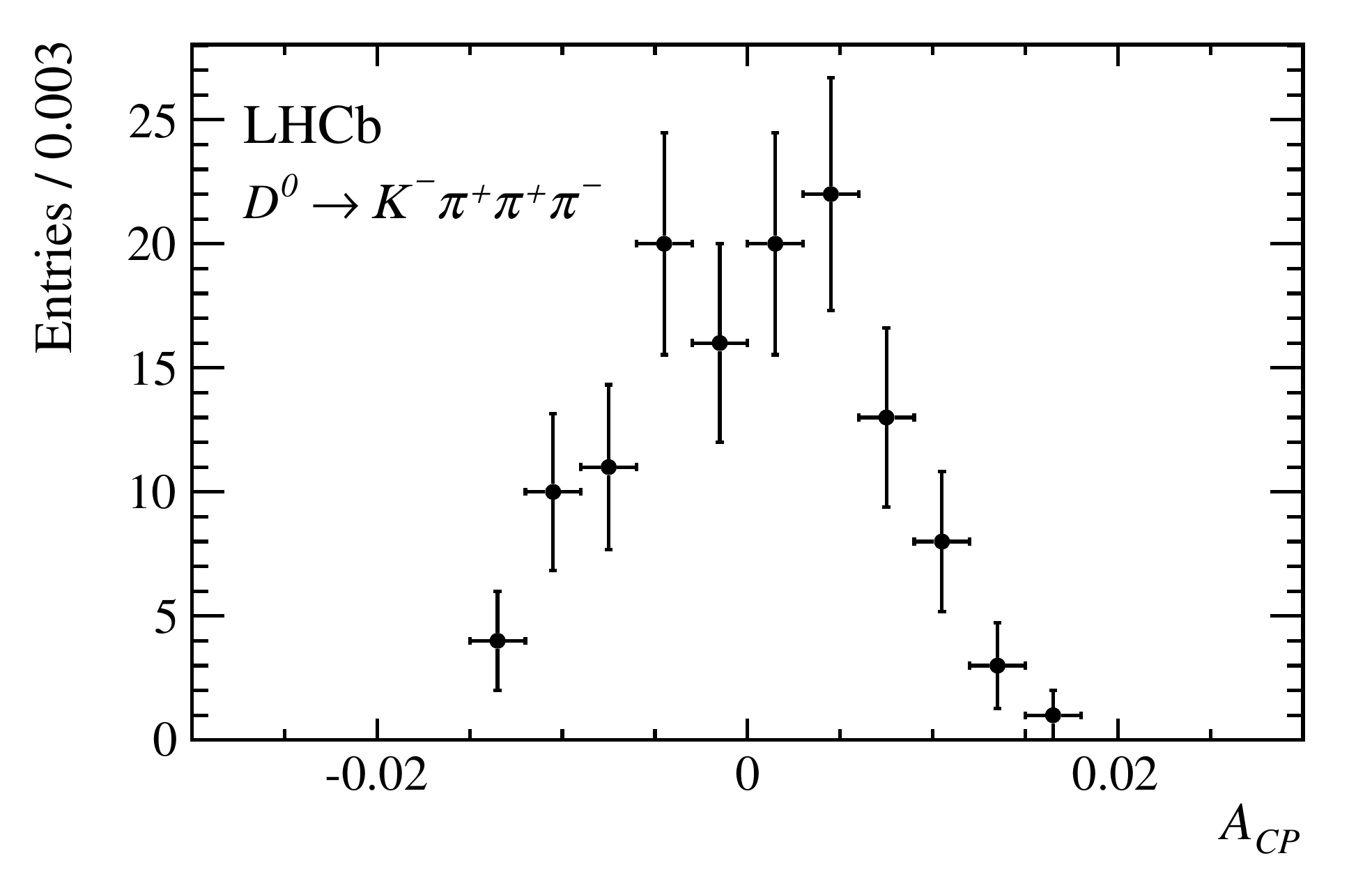}%
       \makebox[0cm][r]{\raisebox{0.19\textheight}[0cm]{\protect\subref{fig:control:RAW}}\hspace{0.04\textwidth}}
     }%

                \caption{\small Distributions of (\protect\subref*{fig:SCP:KKPiPi},\protect\subref*{fig:SCP:FourPi},\protect\subref*{fig:control:SCP})    \SCP  and   (\protect\subref*{fig:RAW:KKPiPi},\protect\subref*{fig:RAW:FourPi},\protect\subref*{fig:control:RAW})   local \CP asymmetry per bin for  (\protect\subref*{fig:SCP:KKPiPi},\protect\subref*{fig:RAW:KKPiPi})  \DKKPiPi decays partitioned with 32 bins, for  (\protect\subref*{fig:SCP:FourPi},\protect\subref*{fig:RAW:FourPi})  \DFourPi decays partitioned with 128 bins, and for (\protect\subref*{fig:control:SCP},\protect\subref*{fig:control:RAW}) the control channel \DKThreePi partitioned with 128 bins.
The points show the data distribution and the solid line is a reference Gaussian distribution corresponding to the no CPV hypothesis. \label{fig:results}}
\end{figure}
\afterpage{\clearpage}

Asymmetries are searched for in the \DKThreePi control channel. The distributions of \SCP and local \CP asymmetry, defined as
\begin{equation*}
 A^{i}_{\CP} = \frac{N_{i}(\Dz) - \alpha N_{i}(\Dzb)}{N_{i}(\Dz) + \alpha N_{i}(\Dzb)},
\end{equation*}
are shown in  \fig{fig:results} for the \DKThreePi control channel. The data set is also studied to identify sources of asymmetry with two alternative partitions and by separating data taken with each magnet polarity. The results, displayed in \tab{tab:PValueControl}, show that no asymmetry is observed in \DKThreePi decays. Furthermore, the data sample is split into 10 time-ordered samples of approximately equal size, for each polarity. The \pvalues under the hypothesis of no asymmetry are uniformly distributed across the data taking period. No evidence for a significant asymmetry in any bin is found.

The \SCP and local \CP asymmetry distributions for \DKKPiPi decays for a partition containing 32 bins and for \DFourPi decays with a partition containing 128 bins are shown in \fig{fig:results}. 
The \pvalues under the hypothesis of no \CP violation for the decays \DKKPiPi and \DFourPi are 9.1\% and 41\%, respectively. The consistency of the results is checked with alternative partitions and the \pvalues are displayed in \tab{tab:PValueResults}. 

The stability of the results is checked for each polarity in 10 approximately equal-sized, time-ordered data samples. The \pvalues are uniformly distributed across the 2011 data taking period 
and are consistent with the no CPV hypothesis. 

\section{Conclusions}

A model-independent search for CPV in $5.7 \times 10^{4}$ \DKKPiPi decays and $3.3 \times 10^{5}$  \DFourPi decays
is presented. 
 The analysis is sensitive to CPV that would arise from a phase difference of $\mathcal{O}$(10$^{\circ}$) or a magnitude difference of $\mathcal{O}$(10$\%$) between
$\decay{\Dz}{\phi \rho^0}$ and $\decay{\Dzb}{ {\phi} {\rho^0}}$ decays for the \DKKPiPi mode  and between $\decay{\Dz}{a_{1}(1260)^{+} \pi^{-}}$ and $\decay{\Dzb}{a_{1}(1260)^{-} \pi^{+}}$ decays for the \DFourPi mode. 
For none of the 32 bins, each with approximately 1800 signal events, is an asymmetry greater than 6.5\% observed for \DKKPiPi decays, 
and for none of the 128 bins, each with approximately 2500 signal events, is an asymmetry greater than 5.5\% observed for \DFourPi decays. 
Assuming \CP conservation, the probabilities to observe local asymmetries across the phase-space of the \Dz meson decay as large or larger than those 
in data for the decays \DKKPiPi and \DFourPi are 9.1\% and 41\%, respectively. All results are consistent with  \CP conservation at the current sensitivity.

\section*{Acknowledgements}

\noindent We express our gratitude to our colleagues in the CERN
accelerator departments for the excellent performance of the LHC. We
thank the technical and administrative staff at the LHCb
institutes. We acknowledge support from CERN and from the national
agencies: CAPES, CNPq, FAPERJ and FINEP (Brazil); NSFC (China);
CNRS/IN2P3 and Region Auvergne (France); BMBF, DFG, HGF and MPG
(Germany); SFI (Ireland); INFN (Italy); FOM and NWO (The Netherlands);
SCSR (Poland); MEN/IFA (Romania); MinES, Rosatom, RFBR and NRC
``Kurchatov Institute'' (Russia); MinECo, XuntaGal and GENCAT (Spain);
SNSF and SER (Switzerland); NAS Ukraine (Ukraine); STFC (United
Kingdom); NSF (USA). We also acknowledge the support received from the
ERC under FP7. The Tier1 computing centres are supported by IN2P3
(France), KIT and BMBF (Germany), INFN (Italy), NWO and SURF (The
Netherlands), PIC (Spain), GridPP (United Kingdom). We are thankful
for the computing resources put at our disposal by Yandex LLC
(Russia), as well as to the communities behind the multiple open
source software packages that we depend on.

%
%
%

\clearpage

\setboolean{inbibliography}{true}
\bibliographystyle{LHCb}
\bibliography{main}

\ifx\mcitethebibliography\mciteundefinedmacro
\PackageError{LHCb.bst}{mciteplus.sty has not been loaded}
{This bibstyle requires the use of the mciteplus package.}\fi
\providecommand{\href}[2]{#2}
\begin{mcitethebibliography}{10}
\mciteSetBstSublistMode{n}
\mciteSetBstMaxWidthForm{subitem}{\alph{mcitesubitemcount})}
\mciteSetBstSublistLabelBeginEnd{\mcitemaxwidthsubitemform\space}
{\relax}{\relax}

\bibitem{Bianco:2003vb}
S.~Bianco, F.~Fabbri, D.~Benson, and I.~Bigi,
  \ifthenelse{\boolean{articletitles}}{{\it {A Cicerone for the physics of
  charm}}, }{}\href{http://dx.doi.org/10.1393/ncr/i2003-10003-1}{Riv.\ Nuovo
  Cim.\  {\bf 26N7} (2003) 1}, \href{http://arxiv.org/abs/hep-ex/0309021}{{\tt
  arXiv:hep-ex/0309021}}\relax
\mciteBstWouldAddEndPuncttrue
\mciteSetBstMidEndSepPunct{\mcitedefaultmidpunct}
{\mcitedefaultendpunct}{\mcitedefaultseppunct}\relax
\EndOfBibitem
\bibitem{Du:2006jc}
D.-S. Du, \ifthenelse{\boolean{articletitles}}{{\it {CP violation for neutral
  charmed meson decays into CP eigenstates}},
  }{}\href{http://dx.doi.org/10.1140/epjc/s10052-007-0242-6}{Eur.\ Phys.\ J.\
  {\bf C50} (2007) 579}, \href{http://arxiv.org/abs/hep-ph/0608313}{{\tt
  arXiv:hep-ph/0608313}}\relax
\mciteBstWouldAddEndPuncttrue
\mciteSetBstMidEndSepPunct{\mcitedefaultmidpunct}
{\mcitedefaultendpunct}{\mcitedefaultseppunct}\relax
\EndOfBibitem
\bibitem{Buccella:2013tya}
F.~Buccella, M.~Lusignoli, A.~Pugliese, and P.~Santorelli,
  \ifthenelse{\boolean{articletitles}}{{\it {CP violation in D meson decays:
  would it be a sign of new physics?}},
  }{}\href{http://arxiv.org/abs/1305.7343}{{\tt arXiv:1305.7343}}\relax
\mciteBstWouldAddEndPuncttrue
\mciteSetBstMidEndSepPunct{\mcitedefaultmidpunct}
{\mcitedefaultendpunct}{\mcitedefaultseppunct}\relax
\EndOfBibitem
\bibitem{Bobrowski:2010xg}
M.~Bobrowski, A.~Lenz, J.~Riedl, and J.~Rohrwild,
  \ifthenelse{\boolean{articletitles}}{{\it {How large can the SM contribution
  to CP violation in $D^0-\overline D^0$ mixing be?}},
  }{}\href{http://dx.doi.org/10.1007/JHEP03(2010)009}{JHEP {\bf 03} (2010)
  009}, \href{http://arxiv.org/abs/1002.4794}{{\tt arXiv:1002.4794}}\relax
\mciteBstWouldAddEndPuncttrue
\mciteSetBstMidEndSepPunct{\mcitedefaultmidpunct}
{\mcitedefaultendpunct}{\mcitedefaultseppunct}\relax
\EndOfBibitem
\bibitem{Grossman:2006jg}
Y.~Grossman, A.~L. Kagan, and Y.~Nir, \ifthenelse{\boolean{articletitles}}{{\it
  {New physics and CP violation in singly Cabibbo suppressed D decays}},
  }{}\href{http://dx.doi.org/10.1103/PhysRevD.75.036008}{Phys.\ Rev.\  {\bf
  D75} (2007) 036008}, \href{http://arxiv.org/abs/hep-ph/0609178}{{\tt
  arXiv:hep-ph/0609178}}\relax
\mciteBstWouldAddEndPuncttrue
\mciteSetBstMidEndSepPunct{\mcitedefaultmidpunct}
{\mcitedefaultendpunct}{\mcitedefaultseppunct}\relax
\EndOfBibitem
\bibitem{Petrov:2010gy}
A.~A. Petrov, \ifthenelse{\boolean{articletitles}}{{\it {Searching for new
  physics with Charm}}, }{}PoS {\bf BEAUTY2009} (2009) 024,
  \href{http://arxiv.org/abs/1003.0906}{{\tt arXiv:1003.0906}}\relax
\mciteBstWouldAddEndPuncttrue
\mciteSetBstMidEndSepPunct{\mcitedefaultmidpunct}
{\mcitedefaultendpunct}{\mcitedefaultseppunct}\relax
\EndOfBibitem
\bibitem{LHCb-PAPER-2013-003}
LHCb collaboration, R.~Aaij {\em et~al.},
  \ifthenelse{\boolean{articletitles}}{{\it {Search for direct \CP violation in
  $D^0 \to h^- h^+$ modes using semileptonic $B$ decays}},
  }{}\href{http://dx.doi.org/10.1016/j.physletb.2013.04.061}{Phys.\ Lett.\
  {\bf B723} (2013) 33}, \href{http://arxiv.org/abs/1303.2614}{{\tt
  arXiv:1303.2614}}\relax
\mciteBstWouldAddEndPuncttrue
\mciteSetBstMidEndSepPunct{\mcitedefaultmidpunct}
{\mcitedefaultendpunct}{\mcitedefaultseppunct}\relax
\EndOfBibitem
\bibitem{LHCb-PAPER-2012-052}
LHCb collaboration, R.~Aaij {\em et~al.},
  \ifthenelse{\boolean{articletitles}}{{\it {Searches for \CP violation in the
  $D^+ \to \phi \pi^+$ and $D_s^+ \to K^0_{\rm S} \pi^+$ decays}},
  }{}\href{http://dx.doi.org/10.1007/JHEP06(2013)112}{JHEP {\bf 06} (2013)
  112}, \href{http://arxiv.org/abs/1303.4906}{{\tt arXiv:1303.4906}}\relax
\mciteBstWouldAddEndPuncttrue
\mciteSetBstMidEndSepPunct{\mcitedefaultmidpunct}
{\mcitedefaultendpunct}{\mcitedefaultseppunct}\relax
\EndOfBibitem
\bibitem{LHCb-PAPER-2011-023}
LHCb collaboration, R.~Aaij {\em et~al.},
  \ifthenelse{\boolean{articletitles}}{{\it {Evidence for \CP violation in
  time-integrated $D^0 \rightarrow h^-h^+$ decay rates}},
  }{}\href{http://dx.doi.org/10.1103/PhysRevLett.108.111602}{Phys.\ Rev.\
  Lett.\  {\bf 108} (2012) 111602}, \href{http://arxiv.org/abs/1112.0938}{{\tt
  arXiv:1112.0938}}\relax
\mciteBstWouldAddEndPuncttrue
\mciteSetBstMidEndSepPunct{\mcitedefaultmidpunct}
{\mcitedefaultendpunct}{\mcitedefaultseppunct}\relax
\EndOfBibitem
\bibitem{Artuso:2012df}
CLEO collaboration, M.~Artuso {\em et~al.},
  \ifthenelse{\boolean{articletitles}}{{\it {Amplitude analysis of $D^0\to
  K^+K^-\pi^+\pi^-$}},
  }{}\href{http://dx.doi.org/10.1103/PhysRevD.85.122002}{Phys.\ Rev.\  {\bf
  D85} (2012) 122002}, \href{http://arxiv.org/abs/1201.5716}{{\tt
  arXiv:1201.5716}}\relax
\mciteBstWouldAddEndPuncttrue
\mciteSetBstMidEndSepPunct{\mcitedefaultmidpunct}
{\mcitedefaultendpunct}{\mcitedefaultseppunct}\relax
\EndOfBibitem
\bibitem{Bediaga:2009tr}
I.~Bediaga {\em et~al.}, \ifthenelse{\boolean{articletitles}}{{\it {On a CP
  anisotropy measurement in the Dalitz plot}},
  }{}\href{http://dx.doi.org/10.1103/PhysRevD.80.096006}{Phys.\ Rev.\  {\bf
  D80} (2009) 096006}, \href{http://arxiv.org/abs/0905.4233}{{\tt
  arXiv:0905.4233}}\relax
\mciteBstWouldAddEndPuncttrue
\mciteSetBstMidEndSepPunct{\mcitedefaultmidpunct}
{\mcitedefaultendpunct}{\mcitedefaultseppunct}\relax
\EndOfBibitem
\bibitem{Aubert:2008yd}
BaBar collaboration, B.~Aubert {\em et~al.},
  \ifthenelse{\boolean{articletitles}}{{\it {Search for \CP violation in
  neutral D meson Cabibbo-suppressed three-body decays}},
  }{}\href{http://dx.doi.org/10.1103/PhysRevD.78.051102}{Phys.\ Rev.\  {\bf
  D78} (2008) 051102}, \href{http://arxiv.org/abs/0802.4035}{{\tt
  arXiv:0802.4035}}\relax
\mciteBstWouldAddEndPuncttrue
\mciteSetBstMidEndSepPunct{\mcitedefaultmidpunct}
{\mcitedefaultendpunct}{\mcitedefaultseppunct}\relax
\EndOfBibitem
\bibitem{LHCb-PAPER-2011-017}
LHCb collaboration, R.~Aaij {\em et~al.},
  \ifthenelse{\boolean{articletitles}}{{\it {Search for \CP violation in $D^{+}
  \to K^{-}K^{+}\pi^{+}$ decays}},
  }{}\href{http://dx.doi.org/10.1103/PhysRevD.84.112008}{Phys.\ Rev.\  {\bf
  D84} (2011) 112008}, \href{http://arxiv.org/abs/1110.3970}{{\tt
  arXiv:1110.3970}}\relax
\mciteBstWouldAddEndPuncttrue
\mciteSetBstMidEndSepPunct{\mcitedefaultmidpunct}
{\mcitedefaultendpunct}{\mcitedefaultseppunct}\relax
\EndOfBibitem
\bibitem{Alves:2008zz}
LHCb collaboration, A.~A. Alves~Jr. {\em et~al.},
  \ifthenelse{\boolean{articletitles}}{{\it {The \lhcb detector at the LHC}},
  }{}\href{http://dx.doi.org/10.1088/1748-0221/3/08/S08005}{JINST {\bf 3}
  (2008) S08005}\relax
\mciteBstWouldAddEndPuncttrue
\mciteSetBstMidEndSepPunct{\mcitedefaultmidpunct}
{\mcitedefaultendpunct}{\mcitedefaultseppunct}\relax
\EndOfBibitem
\bibitem{LHCb-DP-2012-003}
M.~Adinolfi {\em et~al.}, \ifthenelse{\boolean{articletitles}}{{\it
  {Performance of the \lhcb RICH detector at the LHC}},
  }{}\href{http://dx.doi.org/10.1140/epjc/s10052-013-2431-9}{Eur.\ Phys.\ J.\
  {\bf C73} (2013) 2431}, \href{http://arxiv.org/abs/1211.6759}{{\tt
  arXiv:1211.6759}}\relax
\mciteBstWouldAddEndPuncttrue
\mciteSetBstMidEndSepPunct{\mcitedefaultmidpunct}
{\mcitedefaultendpunct}{\mcitedefaultseppunct}\relax
\EndOfBibitem
\bibitem{LHCb-DP-2012-004}
R.~Aaij {\em et~al.}, \ifthenelse{\boolean{articletitles}}{{\it {The \lhcb
  trigger and its performance in 2011}},
  }{}\href{http://dx.doi.org/10.1088/1748-0221/8/04/P04022}{JINST {\bf 8}
  (2013) P04022}, \href{http://arxiv.org/abs/1211.3055}{{\tt
  arXiv:1211.3055}}\relax
\mciteBstWouldAddEndPuncttrue
\mciteSetBstMidEndSepPunct{\mcitedefaultmidpunct}
{\mcitedefaultendpunct}{\mcitedefaultseppunct}\relax
\EndOfBibitem
\bibitem{Hulsbergen:2005pu}
W.~D. Hulsbergen, \ifthenelse{\boolean{articletitles}}{{\it {Decay chain
  fitting with a Kalman filter}},
  }{}\href{http://dx.doi.org/10.1016/j.nima.2005.06.078}{Nucl.\ Instrum.\
  Meth.\  {\bf A552} (2005) 566},
  \href{http://arxiv.org/abs/physics/0503191}{{\tt
  arXiv:physics/0503191}}\relax
\mciteBstWouldAddEndPuncttrue
\mciteSetBstMidEndSepPunct{\mcitedefaultmidpunct}
{\mcitedefaultendpunct}{\mcitedefaultseppunct}\relax
\EndOfBibitem
\bibitem{JohnsonFunction}
N.~L. Johnson, \ifthenelse{\boolean{articletitles}}{{\it Systems of frequency
  curves generated by methods of translation}, }{}Biometrika {\bf 36} (1949)
  149\relax
\mciteBstWouldAddEndPuncttrue
\mciteSetBstMidEndSepPunct{\mcitedefaultmidpunct}
{\mcitedefaultendpunct}{\mcitedefaultseppunct}\relax
\EndOfBibitem
\bibitem{Skwarnicki:1986xj}
T.~Skwarnicki, {\em {A study of the radiative cascade transitions between the
  Upsilon-prime and Upsilon resonances}}, PhD thesis, Institute of Nuclear
  Physics, Krakow, 1986,
  {\href{http://inspirehep.net/record/230779/files/230779.pdf}{DESY-F31-86-02}}\relax
\mciteBstWouldAddEndPuncttrue
\mciteSetBstMidEndSepPunct{\mcitedefaultmidpunct}
{\mcitedefaultendpunct}{\mcitedefaultseppunct}\relax
\EndOfBibitem
\bibitem{Pivk:2004ty}
M.~Pivk and F.~R. Le~Diberder, \ifthenelse{\boolean{articletitles}}{{\it
  {sPlot: a statistical tool to unfold data distributions}},
  }{}\href{http://dx.doi.org/10.1016/j.nima.2005.08.106}{Nucl.\ Instrum.\
  Meth.\  {\bf A555} (2005) 356},
  \href{http://arxiv.org/abs/physics/0402083}{{\tt
  arXiv:physics/0402083}}\relax
\mciteBstWouldAddEndPuncttrue
\mciteSetBstMidEndSepPunct{\mcitedefaultmidpunct}
{\mcitedefaultendpunct}{\mcitedefaultseppunct}\relax
\EndOfBibitem
\bibitem{Link:2007fi}
FOCUS collaboration, J.~Link {\em et~al.},
  \ifthenelse{\boolean{articletitles}}{{\it {Study of the $D^0 \to \pi^{-}
  \pi^{+} \pi^{-} \pi^{+}$ decay}},
  }{}\href{http://dx.doi.org/10.1103/PhysRevD.75.052003}{Phys.\ Rev.\  {\bf
  D75} (2007) 052003}, \href{http://arxiv.org/abs/hep-ex/0701001}{{\tt
  arXiv:hep-ex/0701001}}\relax
\mciteBstWouldAddEndPuncttrue
\mciteSetBstMidEndSepPunct{\mcitedefaultmidpunct}
{\mcitedefaultendpunct}{\mcitedefaultseppunct}\relax
\EndOfBibitem
\bibitem{PDG2012}
Particle Data Group, J.~Beringer {\em et~al.},
  \ifthenelse{\boolean{articletitles}}{{\it {\href{http://pdg.lbl.gov/}{Review
  of particle physics}}},
  }{}\href{http://dx.doi.org/10.1103/PhysRevD.86.010001}{Phys.\ Rev.\  {\bf
  D86} (2012) 010001}\relax
\mciteBstWouldAddEndPuncttrue
\mciteSetBstMidEndSepPunct{\mcitedefaultmidpunct}
{\mcitedefaultendpunct}{\mcitedefaultseppunct}\relax
\EndOfBibitem
\end{mcitethebibliography}

\end{document}